\documentclass[JHEP,12pt]{article}

\usepackage{graphics,color,soul}
\usepackage{graphicx}
\usepackage{amssymb}
\usepackage[export]{adjustbox}
\usepackage{jheppub}
\usepackage{lmodern,mathtools}
\usepackage{caption}
\usepackage{subcaption}
\usepackage[normalem]{ulem}
\usepackage{booktabs}
\usepackage[english]{babel}
\usepackage{amsmath,amssymb,amsbsy,amstext, amsthm, simplewick}
\usepackage{hyperref}
\usepackage{tikz}
\usepackage{physics}
\usetikzlibrary{decorations.markings,decorations.pathmorphing, patterns,shapes}
\usepackage{xcolor}

\usepackage{caption}

\usepackage{amsfonts}
\usepackage{amssymb}
\usepackage{upgreek}
\usepackage{simplewick}
 \usepackage{exscale,relsize}
\usepackage{mathtools}
\usepackage{comment}

\def \p {\partial}
\def \lam {\lambda}

\def\be#1\ee{\begin{align}#1\end{align}}

\def\le{\left}
\def\ri{\right}
\newcommand\ov{\over}
\newcommand{\es}[2] {\begin{equation} \label{#1} \begin{split} #2 \end{split} \end{equation}}

\def\<{\langle}
\def\>{\rangle}
\newcommand\al{{\alpha}}

\newcommand\om{\omega}

\newcommand\ga{{\ensuremath{{\gamma}}}}
\newcommand\Ga{{\ensuremath{{\Gamma}}}}
\newcommand\de{{\ensuremath{{\delta}}}}

\newcommand\betaJ{\beta \mathcal{J}}

\begin{document}

\title{On the temperature dependence of quasinormal modes in SYK and holography
}

\author[a]{Matthew Dodelson,}
\emailAdd{mdodelson@fas.harvard.edu}
\author[b]{Om Gupta,}
 \emailAdd{om.gupta@maths.ox.ac.uk}
\author[b]{M\'ark Mezei,}
 \emailAdd{mark.mezei@maths.ox.ac.uk}
\author[a]{Diandian Wang}
\emailAdd{diandianwang@fas.harvard.edu}

\affiliation[a]{Center for the Fundamental Laws of Nature, Harvard University,
\\
17 Oxford Street, Cambridge, MA 02138, United States}
\affiliation[b]{Mathematical Institute, University of Oxford,
\\
Woodstock Road, Oxford, OX2 6GG, United Kingdom}

\abstract{It was recently found that the quasinormal modes (or Ruelle--Pollicott resonances) of the SYK model at infinite temperature form a Christmas tree shape, reminiscent of AdS black holes. We generalise this computation to finite temperature, allowing us to continuously connect the infinite temperature results to the low temperature regime dual to JT gravity. We contrast the movement of the quasinormal modes with a few examples: various  AdS black holes, dynamical phase transitions, and the large $p$ SYK chain. We find that the relaxation rate increases monotonically with temperature only at strong coupling, corresponding to the gravitational regime.
Byproducts of our investigations are new results on operator growth that may be of independent interest.
}
\maketitle

\section{Introduction}

Quasinormal modes (QNMs) have played an important role throughout the history of holography~\cite{Maldacena:1997re,Witten:1998qj,Gubser:1998bc}, see~\cite{Berti:2009kk} for a review. There has been a growing interest in them in quantum many-body physics, where they are referred to as Ruelle--Pollicott resonances~\cite{prosen2002ruelle,Mori:2023qbd}. They also play a role in the dynamics of astrophysical black holes, and they have recently been measured in gravitational waves~\cite{LIGOScientific:2025wao}. They also have very rich mathematical properties, e.g., a connection with Seiberg--Witten theory~\cite{Aminov:2020yma}.

There are many distinct notions of thermalisation in physical systems: depending on what observable we are considering, its approach to the thermal value and the time scale at which this happens will be different. The simplest observable is the response function
\es{GR}{
G_R(t)\sim\sum_n d_n \,e^{-i \om_n t}\,,
}
where $\om_n$ are the QNM frequencies that lie in the lower half (complex $\om$) plane.\footnote{QNMs do not exist in quantum systems with an (effectively) finite dimensional Hilbert space, as in those systems the correlator does not decay to zero. They require the system to be in the thermodynamic limit, either through taking the volume or the number of local degrees of freedom $N$ to be infinite.} Note that since $G_R(t)$ is real, complex QNMs occur in pairs related by reflection about the imaginary $\omega$ axis. In frequency space, QNMs correspond to poles in the lower half frequency plane of the Fourier transformed retarded Green's function $G_R(\omega)$~\cite{Son:2002sd}. 
QNMs control the approach of correlation functions to thermal equilibrium.
In this paper, we investigate the temperature dependence of the QNMs. 

In~\cite{Dodelson:2024atp}, it was found that the QNMs of the SYK model~\cite{KitaevTalks,Maldacena:2016hyu,Polchinski:2016xgd} at infinite temperature show a strikingly similar pattern to the Christmas tree shape of QNMs of higher-dimensional AdS black holes. This result fits into the recent larger direction of research that aimed at uncovering the shared physics between generic large-$N$ chaotic theories and black holes~\cite{Festuccia:2005pi,
Festuccia:2006sa,
Saad:2019lba,Leutheusser:2021qhd,
Leutheusser:2021frk,
Dodelson:2022eiz,Dodelson:2023vrw,
Ouseph:2023juq,
Grozdanov:2024wgo,Gesteau:2024rpt}. At infinite temperature, the SYK model is perhaps best thought of as a  stringy black hole. When the temperature is taken to be small, the SYK model does have a proper gravity dual, namely JT gravity coupled to matter~\cite{KitaevTalks,Maldacena:2016upp,Jensen:2016pah,Engelsoy:2016xyb}. However, the QNMs of this simple low-dimensional theory lie on the imaginary axis, which in particular do not form a Christmas tree. It is therefore interesting to investigate how the QNMs change with temperature and how the two patterns at infinite and at low temperature morph into each other. 

In Section~\ref{sec:counterexamples_SYK}, we find that the movement of QNMs in the SYK model is complicated at intermediate temperature.  There are two mechanisms by which QNMs continuously transition from a Christmas tree structure to the purely imaginary JT gravity spectrum as we change the temperature: a conjugate pair of QNMs on the Christmas tree collide at a point on the imaginary axis and then separate into two QNMs on the axis, and QNMs already on the imaginary axis move towards the real axis, scaling as $1/\beta$ as $\beta\to \infty$.  Meanwhile, generic QNMs in the complex plane stay in place, leading to the latter disappearing from the spectrum in the very low temperature regime. We indeed observe both of these dynamics in the (somewhat complicated) movement of the QNMs at intermediate temperature  in Figures~\ref{fig:betascan},~\ref{fig:QNMflow}. In the low temperature, strongly coupled, gravity regime the movement of the QNMs becomes orderly, and they all move towards the real frequency axis as temperature is lowered. 

This orderly movement is intuitive:
one may expect that heating up a system should speed up thermalisation: 
\es{hotfast}{
\frac{d|\text{Im }\omega_n|}{dT}\stackrel{?}{>}0\,.
}
This would follow from an excitation having more energetically allowed decay channels.\footnote{This rough intuition only applies to the first QNM (with smallest imaginary part), since it dominates the physics of generic correlators. Even if we chose a finely tuned operator ${\cal O}$ such that $d_1=0$ in $G_R(t)$ at one temperature and hence $\om_2$ would dominate, $d_1$ may become non-zero as we change the temperature.} Indeed in Section~\ref{sec:examples} we analyse AdS black holes, dual to thermal states in large-$N$ gauge theories, and in a variety of limits we show analytically and in other regimes we provide numerical evidence that~\eqref{hotfast} holds. If the motion of the QNMs satisfies the inequality opposite to~\eqref{hotfast}, we will call it anti-monotonic; if it satisfies neither condition, we will call it non-monotonic.

On second thought, however, we know a few counterexamples to~\eqref{hotfast} even in the gravity regime. Consider a finite temperature CFT that exhibits energy diffusion. There are several holographic CFTs of this kind. The dispersion relation of the QNM associated to energy diffusion is (with $k$ held fixed):
\es{DiffDisp}{
\om(k)=-iDk^2+\dots\,, \qquad D={c_D\ov  T}\,,
}
where $c_D$ is a dimensionless constant and the  dependence on $T$ follows from dimensional analysis. We provide a quasiparticle intuition for this result in Section~\ref{sec:hydropole}. Then in Section~\ref{sec:SSB} we review that restoration of a continuous symmetry can lead to the violation of~\eqref{hotfast} due to the topological structure of the movement of QNMs including the Goldstone mode. 

We conclude that if generic QNMs, not related to symmetries and conservation laws, obey~\eqref{hotfast}, it may indicate a holographic gravity dual description. As additional evidence, we analyse the QNMs appearing in the energy density retarded Green's function of the large $p$ SYK chain~\cite{Choi:2020tdj}: their movement is also complicated, and only becomes orderly in the strongly coupled regime.\footnote{It is not known what the gravity dual of the low temperature SYK chain is, but it may be a close relative of near-extremal black branes.}

The main technical innovation that allows us to compute QNMs at finite temperature is setting up the real-time Schwinger-Dyson equations in the SYK model at finite temperature in a way amenable to the  finite temperature generalisation of the algorithm described in~\cite{Parker:2018yvk}. This allows us to recursively generate Taylor series coefficients $\mu_{2n}$ of the Green's functions in the finite $p$ SYK model. Subsequently, we can compute QNMs following methods discussed in~\cite{Dodelson:2024atp,Dodelson:2025rng}. 

In order to obtain results at low temperatures, we made use of Pad\'e approximants to continue the Taylor series generated by our algorithm beyond its radius of convergence. We then find that given only the Taylor series data around infinite temperature, the Pad\'e approximant manages to reorganise itself into a smooth interpolation to the JT gravity answer expected at low temperatures with very high accuracy, reproducing the theoretical value within $1.5\%$.

Byproducts of our investigations include some new results on operator growth that may be of independent interest; we collect these in Section~\ref{sec:opgrowth}. We can use  the data contained in the finite temperature Green's functions to compute two measures of operator growth:
the $\lambda_L$ Lyapunov exponent~\cite{kitaev,Maldacena:2015waa} and the Krylov exponent $\alpha=\pi/\beta_0$. $\beta_0$ is an effective inverse temperature that defines a strip of analyticity of the Wightman function on the complex $t$ plane, and hence obeys $\beta\leq\beta_0$.  In~\cite{Parker:2018yvk}, a bound was conjectured relating the two
\es{intro_bound}{
\lambda_L\leq \frac{2\pi}{\beta_0}\,.
}
Note that this is a stronger inequality than the chaos bound $\lambda_L\leq2\pi/\beta$ proven in~\cite{Maldacena:2015waa}. For the SYK model the inequality~\eqref{intro_bound} was proven in~\cite{Gu:2021xaj}.
 Here we provide new evidence that the bound is obeyed as a strict inequality
in the high temperature regime of the $p=4$ SYK model.\footnote{Our methods lose accuracy around $\beta{\cal J}\approx 4$. At lower intermediate temperature Figure 9 in Appendix C of~\cite{Gu:2021xaj} provides convincing numerical evidence for the bound.} This is in contrast to the large $p$ model, where $\lambda_L= 2\pi/\beta_0$~\cite{Parker:2018yvk}. We also conjecture a momentum dependent refinement of~\eqref{intro_bound} and show that it is obeyed in the large $p$ SYK chain as a  strict inequality.

We end with a brief discussion in Section~\ref{sec:discussion} and include two appendices with technical details.

\section{QNMs in the SYK model}\label{sec:counterexamples_SYK}

The SYK model shows a lot of qualitative similarity to AdS black holes, including its chaotic operator growth and the Christmas tree shape of its quasinormal modes. 
In this section, we will look at the trajectory of QNMs in the finite $p$ SYK model as a function of temperature. We will find that the movement is complicated at high temperature and comprises of collisions of pairs of QNMs on the imaginary axis, but eventually settles into a monotonic movement towards the real frequency axis in the low temperature, gravity regime. We can also see how the Christmas tree at infinite temperature morphs into the purely imaginary spectrum of JT gravity. 

Before doing so, we review tools from Krylov space necessary for our computation, and describe the finite temperature generalisation of the algorithm described in~\cite{Parker:2018yvk}. We will also remark on the technical details of our algorithm and computation of QNMs --- in particular, on the role of Pad\'e approximants in obtaining results for low temperature.
\subsection{Generalised moments at finite temperature} 
The basic objects of interest are the two-sided correlator $C(t)$ and the retarded Green's function $G_R(t)$, defined by
\begin{align}
C(t)&=\frac{\text{Tr}\left(e^{-\beta H}\mathcal{O}\!\left(t-\frac{i\beta}{2}\right)\mathcal{O}(0)\right)}{\text{Tr }e^{-\beta H}},\\
G_R(t)&=-\frac{i}{2}\frac{\text{Tr}\left(e^{-\beta H}\{\mathcal{O}(t),\mathcal{O}(0)\}\right)}{\text{Tr }e^{-\beta H}}\, \theta(t)\,,
\end{align}
where $\mathcal{O}(t)$ is a Hermitian operator. In frequency space, these two functions are related by
\es{CGRrelation}{
C(\omega)=-\frac{2\text{Im }G_R(\omega)}{\cosh\left(\frac{\beta \omega}{2}\right)}\,,
}
where we have taken the operator $\mathcal{O}$ to be fermionic. In Fourier space the singularities of  $G_R(\om)$ are  the QNMs that lie in the lower half plane due to causality and come in complex conjugate pairs $\omega_n,\omega_n^*$. $C(\om)$ instead has singularities at $\omega_n,\omega_n^*,-\omega_n,-\omega_n^*$~\cite{Dodelson:2023vrw}.

Now let us consider a purely fermionic theory, like a spin system or SYK, with a locally finite dimensional Hilbert space. Then correlation functions at infinite temperature are well-defined, and the two-sided correlator admits a double series expansion in $\beta$ and $t$, 
\begin{align}\label{momentexpansion}
C(t)=\sum_{n,k=0}^{\infty}\frac{\mu_{2n,2k}}{(2n)!(2k)!}(i\mathcal{J}t)^{2n}\left(i\betaJ\right)^{2k}\,.
\end{align}
The coefficients $\mu_{2n,0}$ are the moments of the correlator at infinite temperature. The universal operator growth hypothesis is a statement about the large $n$ behaviour of these moments~\cite{Parker:2018yvk}, 
\begin{align}\label{uogh}
\mu_{2n,0}\sim \left(\frac{4n}{e\beta_0}\right)^{2n}\,,\hspace{10 mm}n\to \infty\,.
\end{align}
This asymptotic behaviour implies that $C(t)$ has a finite radius of convergence $|t|<\beta_0/2$, with a singularity on the imaginary axis at $t=\pm i\beta_0/2$.\\
\indent The sum over $k$ is much less understood than the sum over $n$, and we are not aware of any results for the large $k$ asymptotics of the generalised moments $\mu_{2n,2k}$. As we will see shortly, this question has a definite (and similar) answer in the case of the SYK model.
\subsection{Moments in SYK} \label{sec:scheme}
Let us specialise to the SYK model, with the operator $\mathcal{O}=\sqrt{2}\psi_1$. Our starting point will be the real time Schwinger-Dyson equations at finite temperature. We will merely state the result in the main text and refer the reader to Appendix~\ref{sec:finiteT} for the derivation,
\es{sdequations1}{
    \omega G_R(\omega)& = 1 + \frac{2 \mathcal{J}^2}{p} G_R(\omega) \tilde{\Sigma}(\omega) \,,\\
    \tilde{\Sigma}(\omega)& = \int \frac{d \omega'}{2 \pi} \int d t\, \frac{\cosh \left(\frac{\beta \omega'}{2}\right)e^{i\omega' t}}{\omega-\omega'} \, C(t)^{p-1}\,.
}
Here, $\tilde{\Sigma}(\omega) \equiv \Sigma(-i \omega)$ is defined to be the analytic continuation of the Euclidean space quantity defined in~\eqref{eq:euclidean}, and the frequency $\omega$ is understood to be slightly above the real axis.

We will now describe the recursive scheme to determine the moments at finite temperature. To get a closed set of three equations for three unknown functions, the Schwinger-Dyson equations are supplemented by the relation between $G_R$ and $C$ given in~\eqref{CGRrelation}. At each order in the recursion, $C(t)$ is given by a polynomial in $t$, and $G_R(\omega)$ is given by a polynomial in $1/\omega$. When going from $G_R(\omega)$ to $C(t)$, we encounter integrals of the form
\begin{align}\label{omegatot} 
-\int \frac{d\omega}{2\pi} \frac{e^{-i\omega t}}{\cosh\left(\frac{\beta \omega}{2}\right)}\,\text{Im}\left(\frac{1}{(\omega+i\epsilon)^n}\right)=\frac{1}{(n-1)!}\partial_\omega^{n-1}\left(\frac{e^{-i\omega t}}{ \cosh\left(\frac{\beta \omega}{2}\right)}\right)\big|_{\omega=0}\,,
\end{align}
where the right hand side is a polynomial in $t$.
The step of going from $C(t)$ to $\tilde{\Sigma}(\omega)$ involves integrals of the form
\begin{align}\label{ttoomega}
 \int \frac{d \omega'}{2 \pi} \int d t\, \frac{\cosh \left(\frac{\beta \omega'}{2}\right)e^{i\omega' t}}{\omega-\omega'} \, t^n=(-i)^n\partial_{\omega'}^{n}\left(\frac{\cosh \left(\frac{\beta \omega'}{2}\right)}{\omega-\omega'} \right)|_{\omega'=0}\,,
\end{align}
where the right hand side is a polynomial in $1/\om$.
The full procedure is as follows:
\begin{enumerate}
    \item Set $G_{R,0}(\omega) = \omega^{-1}$.
    \item In the $j$th step of the iteration, compute $C_{j}(t)$ from $G_{R,j}(\omega)$ by using the relation~\eqref{CGRrelation} and by replacing $\omega^{-n}$ in the relevant integral with the right hand side of~\eqref{omegatot}.
    \item Compute $\Sigma_j(\omega)$ from $C_j(t)$ by using the second equation in~\eqref{sdequations1} and by replacing $t^n$ in the relevant integral with the right hand side of~\eqref{ttoomega}.
    \item Set $G_{R,j+1}(\omega) =\omega^{-1}(1 + 2 p^{-1}\mathcal{J}^2G_{R,j}(\omega)\tilde{\Sigma}_j(\omega) )$ up to order $\omega^{-2j-3}$.
\end{enumerate}

\indent After running this algorithm, we can read off the moments $\mu_{2n,2k}$ from the Taylor expansion of $C_j(t)$. At large $n$ and fixed $k$, we observe the standard asymptotic behaviour~\eqref{uogh}. Interestingly, the large $k$ asymptotics at fixed $n$ take a similar form,
\begin{align}\label{largekasy}
\mu_{2n,2k}\sim\left(\frac{4k}{e\tilde{\beta}_0}\right)^{2k}\,,\hspace{10 mm}k\to \infty\,.
\end{align}
At a fixed time $t$, the asymptotics~\eqref{largekasy} then suggest a simple interpretation: the Taylor series in $\betaJ$ for $C(t)$ is convergent with a radius of convergence $\tilde{\beta}_0/2$ and singularities at $\beta = \pm \tilde{\beta}_0/2$. Indeed, our numerics for the case $p = 4$ confirm this expectation and we find that $\tilde{\beta}_0 \mathcal{J}/2 \approx 1.4$. It would be interesting to explore the dependence of $\tilde{\beta}_0$ on $p$ by running the algorithm for different $p$ but we restrict attention to the case $p =4$ in this work. From now on, we set $\mathcal{J} =1$ so that we will use $\beta$ and $\betaJ$ interchangeably. 

\subsection{From moments to QNMs}
In this section, we will briefly review the two methods~\cite{Dodelson:2024atp,Dodelson:2025rng} that allow us to compute QNMs given the finite temperature moment generating algorithm of the previous section. We find it is necessary to use both the methods we describe here for reasons discussed in the next section.

The first approach is that of~\cite{Dodelson:2024atp}, which fits a sum of exponentials to the Taylor expansion generated by the algorithm, i.e., 
\begin{equation*}
    \sum_{n=0}^{K}\mu_{2n} \frac{(it)^{2n}}{(2n)!} = C(t) \overset{!}{=} \sum_{n = 0}^{S-1} d_n e^{-i \omega_n t} \,,
\end{equation*}
where $S$ denotes a cutoff chosen by hand. $C(t)$ is then sampled at $2S$ points for $0< t< t_{\text{max}}$ where $t_{\text{max}}$ is another value chosen by hand such that $t_{\text{max}} < \beta_0/2$. This leads to a linear algebra problem for the $2S$ unknowns $d_n$ and $\omega_n$ which can be solved numerically. As the cutoff $S \to \infty$, we interpret $\omega_n$ to be the QNMs of the system we are interested in. For further details about the scheme, we refer readers to Section 3 of~\cite{Dodelson:2024atp}. 

The second approach is that of \cite{Dodelson:2025rng} where the computation of the retarded Green's function is rewritten as a discrete scattering problem on a black hole-like background using Krylov space tools. More precisely, one can construct a ``bulk" retarded Green's function that satisfies the following difference equation,
\begin{equation}
    i \omega G_R(\omega, n, n') + b_{n+1} G_R(\omega, n+1, n') - b_n G_R(\omega, n-1, n') = i \delta_{n,n'}\,. \label{eq:bulkGreen}
\end{equation}
Here, $b_n$ are the Lanczos coefficients which contain identical data to the moments $\mu_{2n}$. One transforms between the two using the following recursion relation~\cite{viswanath1994recursion}
\begin{equation}
    \begin{aligned}
b_n &= \sqrt{M_{2n}^{(n)}}\,, \\
M_{2k}^{(m)} &= \frac{M_{2k}^{(m-1)}}{b_{m-1}^2} - \frac{M_{2k-2}^{(m-2)}}{b_{m-2}^2}\,, \quad k = m, \dots, n\,, \\
M_{2k}^{(0)} &= \mu_{2k}\,, \quad b_{-1} = b_0 := 1\,, \quad M_{2k}^{(-1)} := 0\,.
\end{aligned} \label{eq:lanczos}
\end{equation}
Following Section 2.2 in~\cite{Dodelson:2025rng}, interpreting $n$ to be a continuous parameter, one can rewrite~\eqref{eq:bulkGreen} as a Schr\"odinger-like wave equation on a black hole background, and in the process, make the identification
\begin{equation}
    n \sim  \frac{1}{\sqrt{1-\frac{r_s}{r}}} \,, \quad \quad r\to r_s\,.
\end{equation}
Hence, this suggests $n \to \infty$ is to be thought of as the near-horizon region, and $n\to 0$ as the boundary of the bulk. 

To solve the discrete scattering problem~\eqref{eq:bulkGreen}, one takes inspiration from~\cite{Festuccia:2005pi,Festuccia:2008zx} to define the linearly independent solutions at the horizon to be
\begin{equation}
    h^R_\omega(n) \sim (-1)^nn^{-\frac{1}{2} + \frac{i \omega \beta_0}{2\pi}}\,, \quad \quad
h^A_\omega(n) \sim n^{-\frac{1}{2} - \frac{i \omega\beta_0}{2\pi}}\,, \quad \quad
n \to \infty\,.
\end{equation}
This allows one to immediately write down a solution to $G_R(\omega,n,n')$. The boundary Green's function $G_R(\omega)$ is then obtained by extrapolation to $n=n'=0$, and the QNMs are found as zeroes of the denominator.  
\subsection{Details of the computation in $p =4$ SYK}
In describing the $p =4$ spectrum at infinite temperature,~\cite{Dodelson:2024atp} computed $\sim2000$ moments $\mu_{2n}$. Adding temperature dependence makes the recursion significantly more intensive so that we manage to compute $K = 116$ moments $\mu_{2n}$, where each of the $\mu_{2n}$ is a power series in $\beta$.\footnote{While one can generate the Taylor series in $\beta$ to arbitrary order at each $n$ given the equations~\eqref{omegatot} and~\eqref{ttoomega} and obtain the $\beta$ dependence exactly, in practice, one needs to work with a cutoff, i.e., the maximum order of $\beta$ that one retains in computations. In our case, we worked with a cutoff at the order $\beta^{150}$. }

The second approach described in the previous section converges to the Green's function $G_R(\omega)$ rapidly in $n$; for example, one can compute the first (pair of) QNM(s) to high precision using just the first four Lanczos coefficients $b_n$. However, a drawback of this method is that convergence is guaranteed only in a strip of the complex frequency plane below the real axis. For $\betaJ$ close to 0, this strip is sufficient to obtain the first two ``levels" of QNMs in the complex plane. However, in the same regime, we find that the method of exponential fitting allows us to plot the first four levels to high accuracy. Hence, we use this method in the high/moderate temperature regime. 

Beyond $\betaJ\sim 5$, our methods are able to approximate reliably only the first six moments $\mu_{2n}$. On the other hand, the method of exponential fitting requires precise knowledge of the first $\sim$20 moments to compute the first level of QNMs to high precision. Hence, we make use of the second method beyond $\betaJ\sim 5$, which allows us to track the leading QNM to $\betaJ \sim 100$. 

In Section~\ref{sec:scheme}, we remarked that the radius of convergence of the power series in $\beta$ is $\tilde{\beta}_0/2 \approx 1.4$ --- the reason for which was singularities at $\beta = \pm\tilde{\beta}_0/2$. However, we have claimed that our methods allow us to compute QNMs up to $\betaJ \sim 100$. To do so, we made use of Pad\'e approximants. A Pad\'e approximant of order $[N,M]$ is a rational function whose numerator and denominator are power series of orders $N$ and $M$ respectively, and whose coefficients are fixed by matching the Taylor expansion of the approximant to Taylor series data (possibly) around several expansion points. Since singularities can be well represented by zeroes of the denominator, Pad\'e approximants represent one way to continue a function beyond its radius of convergence. See~\cite{Kotikov:2003fb,Sen:2013oza,Banks:2013nga,Heller:2015dha} for examples of the use of Pad\'e approximants in the high energy literature. 

Here, we would like to interpolate between the $\beta =0$ and $\beta \to \infty$ results for $C(t)$. Hence, we chose to Pad\'e approximate each of the moments $\mu_{2n}$. A natural choice for the Taylor series data would be the expansion generated by the algorithm around $\beta = 0$, and the JT gravity/conformal answer around $\beta \to \infty.$ For $p =4$, the conformal answer gives~\cite{Maldacena:2016hyu,Polchinski:2016xgd} 
\begin{equation} \label{eq:conformaltp}
    C(t) = \frac{1}{\sqrt{\cosh \frac{\pi t}{\beta}}} = \sum_{n} \mu'_{2n} \frac{(it)^{2n}}{(2n)!\beta^{2n}} \,,
\end{equation} 
where $\mu'_{2n}=\{1,\pi^2/2,7\pi^4/4,\dots\}$ are straightforward to read off.
We can then match the Taylor series data at $\beta \to \infty$ as $\mu_{2n} \to \mu'_{2n}/{\beta^{2n}}$. 

Hence, to Pad\'e approximate the moments, our initial approach was as follows: construct an approximant of order $[N,N+2n]$ for the moment $\mu_{2n}$ where $a_j,b_j$ represent coefficients of the power series in the numerator and denominator, respectively. Then the ratio $a_N/b_{N+2n} = \mu'_{2n}$ is fixed by the conformal behaviour, and the rest of the coefficients are matched with the expansion at $\beta =0$. 

\begin{figure}
    \centering
    \includegraphics[width=0.7\linewidth]{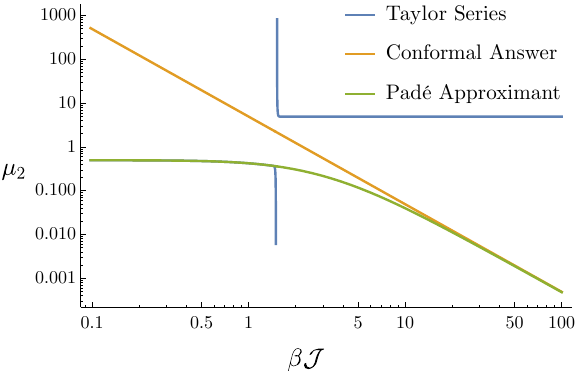}
    \caption{We plot three different representations of $\mu_2$: the Taylor expansion generated by the algorithm (blue), the conformal/JT gravity limit (orange), and the Pad\'e approximant supplied with \emph{only} the Taylor series around $\betaJ =0$ (green). As discussed in the main text, we note that the Pad\'e manages to reorganise the Taylor series around $\beta = 0$ into the asymptotic conformal behaviour.}
    \label{fig:comparison}
\end{figure}

In practice, however, we found it was \emph{unnecessary} to supply Taylor series data about the conformal behaviour. More precisely, we find that a Pad\'e approximant of order $[N,N+2n]$ for the moment $\mu_{2n}$ supplied with \emph{only} the expansion around $\beta = 0$ behaves like $\mu'_{2n}/\beta^{2n}$ as $\beta \to \infty$, with the ratio $a_N/b_{N+2n}$ monotonically approaching $\mu'_{2n}$ as $N$ increases, i.e.,~as one uses more terms in the expansion around $\beta = 0$.\footnote{In particular, we find that $a_N/b_{N+2}$ is within $1.5\%$ of $\mu'_2=\pi^2/2$ for the maximum value of $N$ we use.} In conclusion, we find that the Pad\'e approximant manages to reorganise the Taylor series around $\beta = 0$ into the asymptotic conformal behaviour, \emph{given only the data around $\beta = 0$}. We plot an instance of such an approximant for $\mu_2$ in Figure~\ref{fig:comparison} where we note the Pad\'e interpolates between the two regimes smoothly.

This allows us to extend our approach beyond $\betaJ \approx 1.4$. An additional point of note is that as in Appendix B of~\cite{Dodelson:2025jff}, we found it helpful to also Pad\'e approximate  $C(t)$ to extend the correlator to all $t$, and then as in~\cite{Dodelson:2025jff}, we find the optimal value of $t_{\text{max}}$ to be $t_{\text{max}}\sim 2\beta_0$. Hence, first we Pad\'e approximate each of the moments $\mu_{2n}$ in $\beta$ and \emph{then}, Pad\'e approximate the correlator $C(t)$ in $t$.
\subsection{The spectrum} \label{sec:spectrum} 

We are now ready to describe the spectrum of QNMs. Before looking at the finite $p$ case, let us briefly recall what the spectrum looks like at strong coupling, and large $p$.
SYK at strong coupling has emergent conformal symmetry in which case the retarded fermion two-point function is given by (see~\eqref{eq:conformaltp})
\begin{equation}\label{eq:CQMcorr}
    G_R(t) \sim \left(\frac{1}{\sinh \frac{\pi t}{\beta}}\right)^{2/p} \,.
\end{equation}
The Fourier transform $G_R(\omega) \sim \Gamma\left(\tfrac{1}{p}-\tfrac{i\beta \omega}{2\pi}\right)$ so that the QNMs are fixed as
\begin{equation}\label{eq:JTQNMs}
    \omega_n = -\frac{2\pi i}{\beta} \left(\frac{1}{p}+n\right)\,, \quad \quad n= 0,1,2,\dots\,.
\end{equation}
This result agrees with the classical limit of the QNMs of JT gravity coupled to a massive fermion (and using the well-known relation between bulk mass and boundary conformal dimension). Putting it in a simpler way,~\eqref{eq:CQMcorr} is just the boundary-to-boundary two-point function of a fermion on rigid AdS$_2$. While it is known how to incorporate all-loop graviton corrections to this answer, we do not need to consider these, since in this paper we are working in the strict large $N$ limit dual to classical gravity.
From~\eqref{eq:JTQNMs} we immediately see that in the gravity regime the inequality~\eqref{hotfast} is obeyed.

In the large $p$ limit, it is possible to interpolate analytically between strong and weak coupling. This interpolation is controlled by the dimensionless parameter $v \in [0,1)$ given by
\begin{equation}
    \beta \mathcal{J} = \frac{\pi v}{\cos \frac{\pi v}{2}}\,. \label{eq:veq}
\end{equation}
 As before, we set $\mathcal{J} = 1$ so that varying $v$ (or $\beta \mathcal{J}$) is akin to changing the temperature, or equivalently the SYK coupling strength. 

The retarded Green's function is found to be~\cite{Maldacena:2016hyu, Tarnopolsky:2018env}
\begin{equation}
   G_R(t) \sim \left[\left(\frac{1}{\cos \left[\pi v\left(\frac{1}{2} - \frac{i t}{\beta}\right)\right]} \right)^{2/p} - \left(\frac{1}{\cos \left[\pi v\left(\frac{1}{2} + \frac{i t}{\beta}\right)\right]} \right)^{2/p}\right]\,. \label{eq:largeqtwo}
\end{equation}
The Fourier transform proceeds similarly, leading to the QNMs
\begin{equation}
    \omega_n = -2\pi i\frac{v}{\beta} \left(\frac{1}{p}+n\right)\,, \quad \quad n= 0,1,2,\dots\,.
\end{equation}
From~\eqref{eq:veq} we can rewrite the temperature dependence of these QNMs to be $\cos \frac{\pi v}{2}$, which is monotonically decreasing on $v\in(0,1)$ and gives that the inequality~\eqref{hotfast} is obeyed.

\begin{figure}
    \centering
    \includegraphics[width=0.32\linewidth]{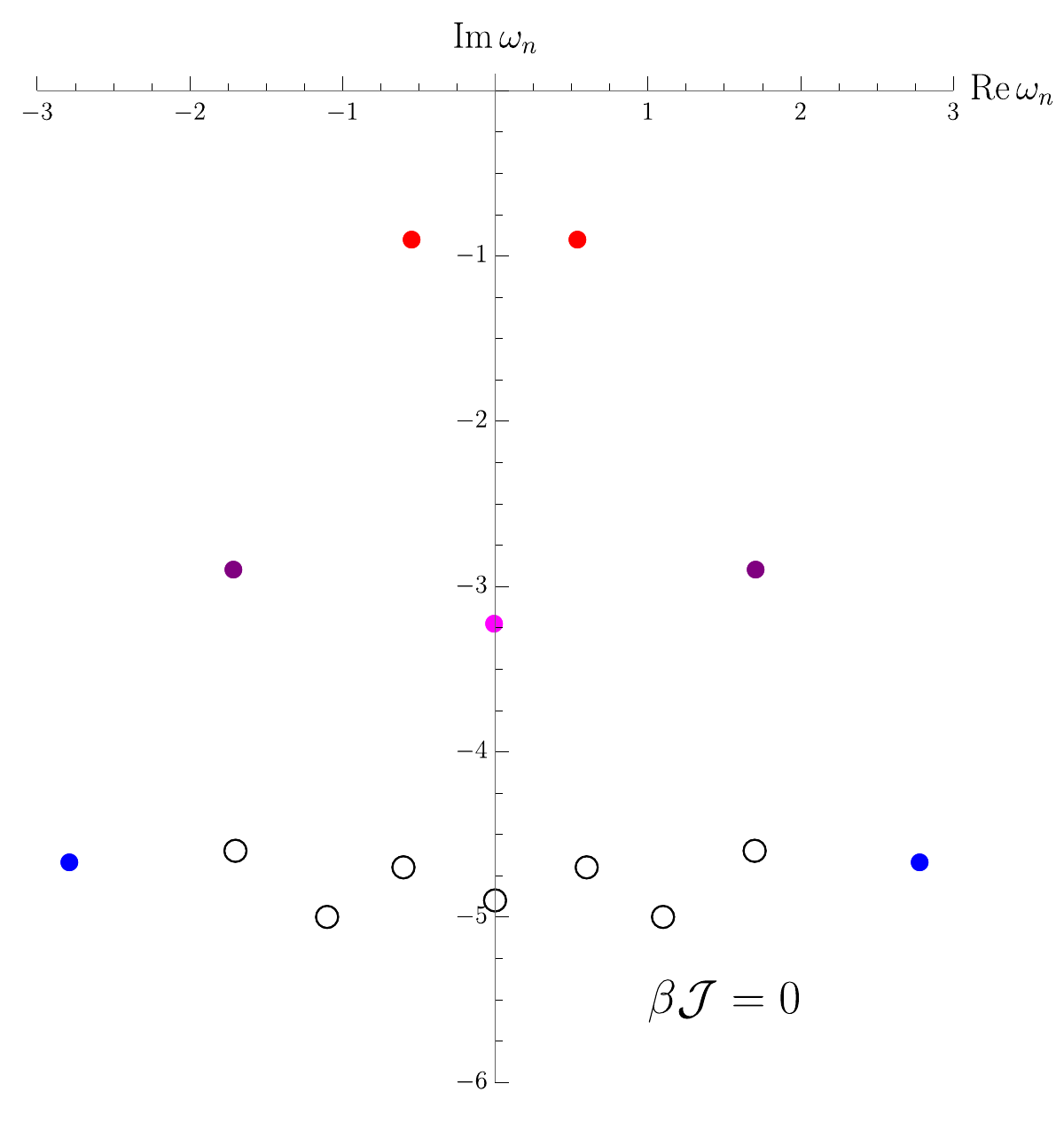}
    \includegraphics[width=0.32\linewidth]{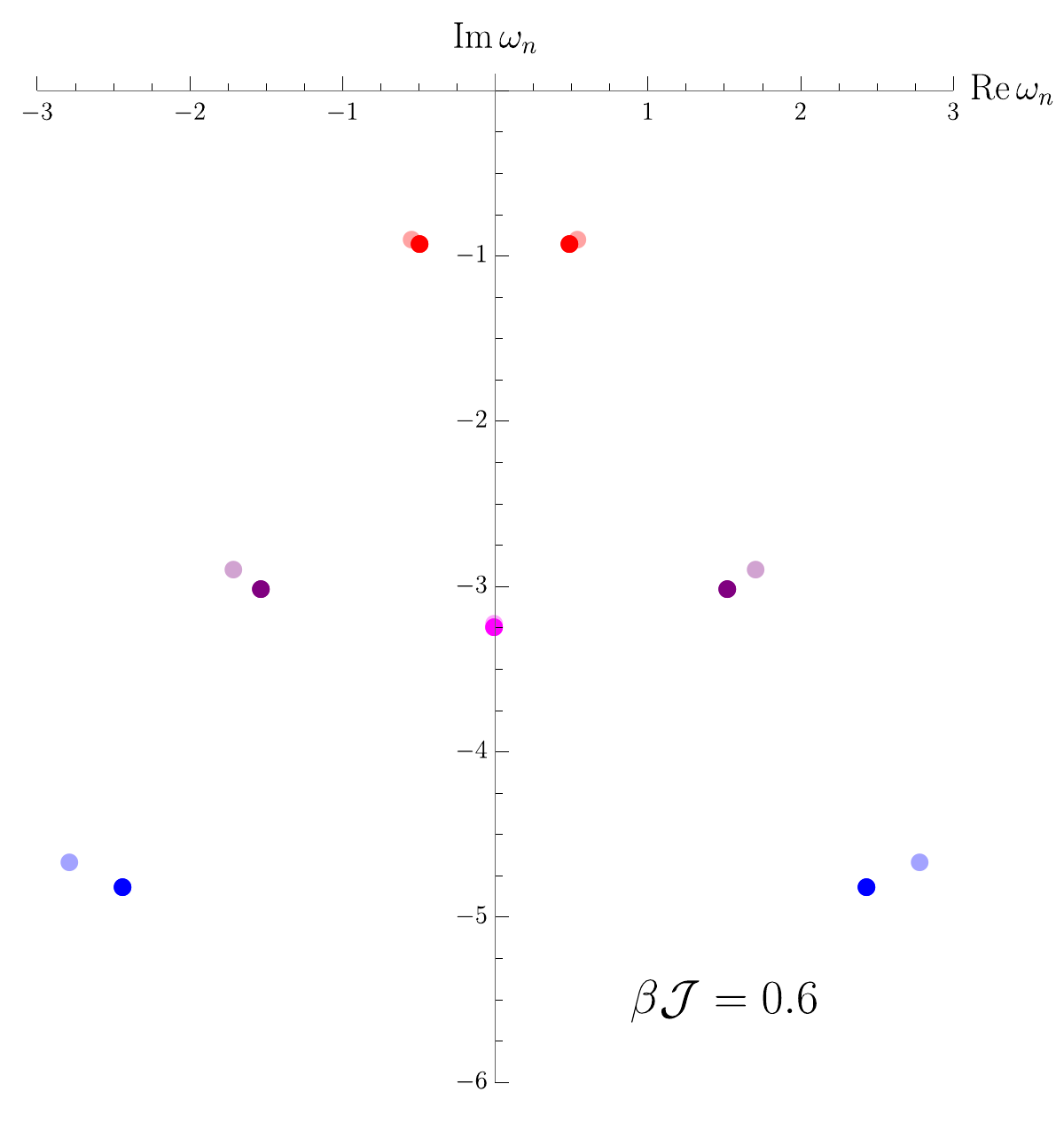}
    \includegraphics[width=0.32\linewidth]{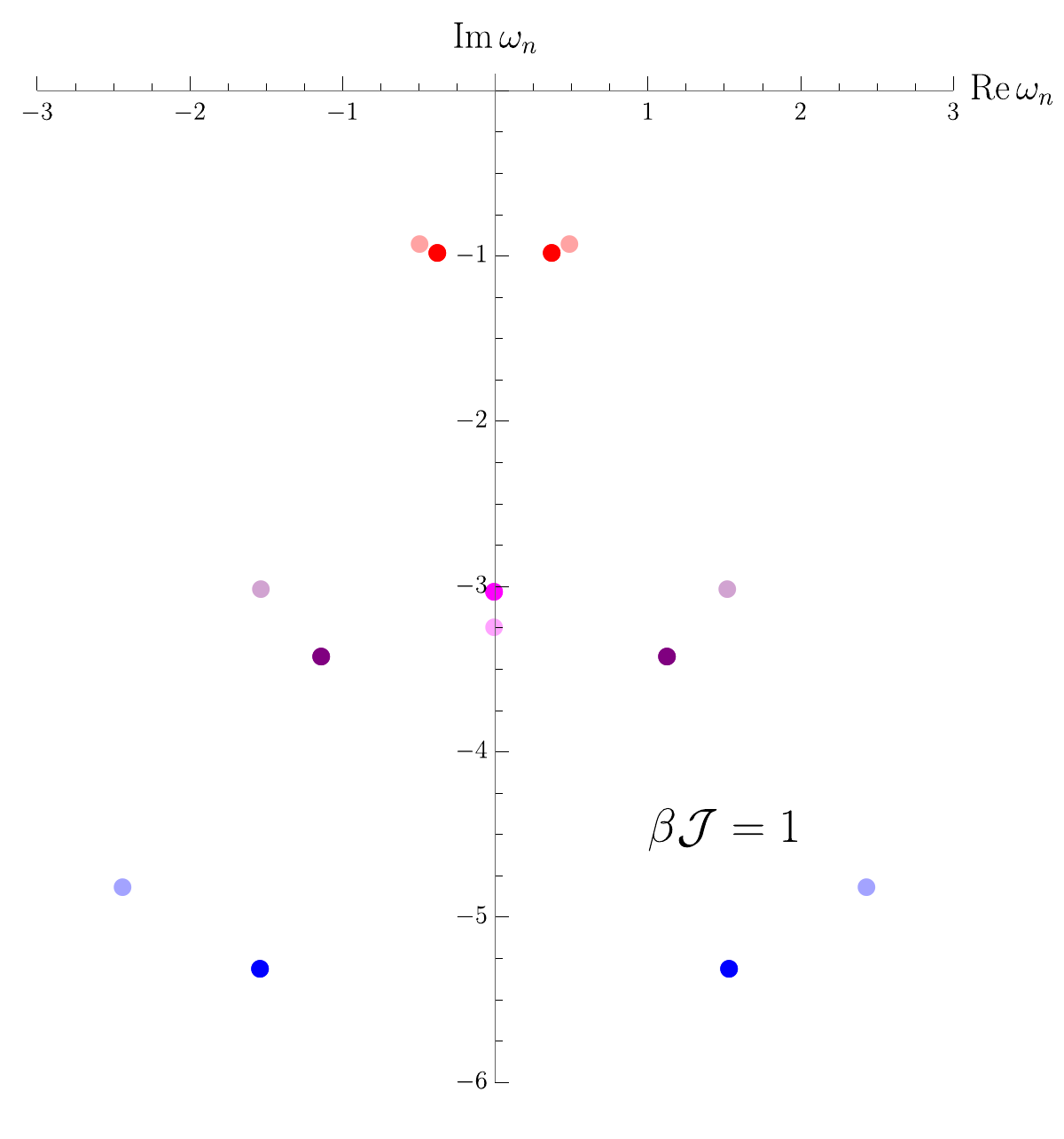}
    \includegraphics[width=0.32\linewidth]{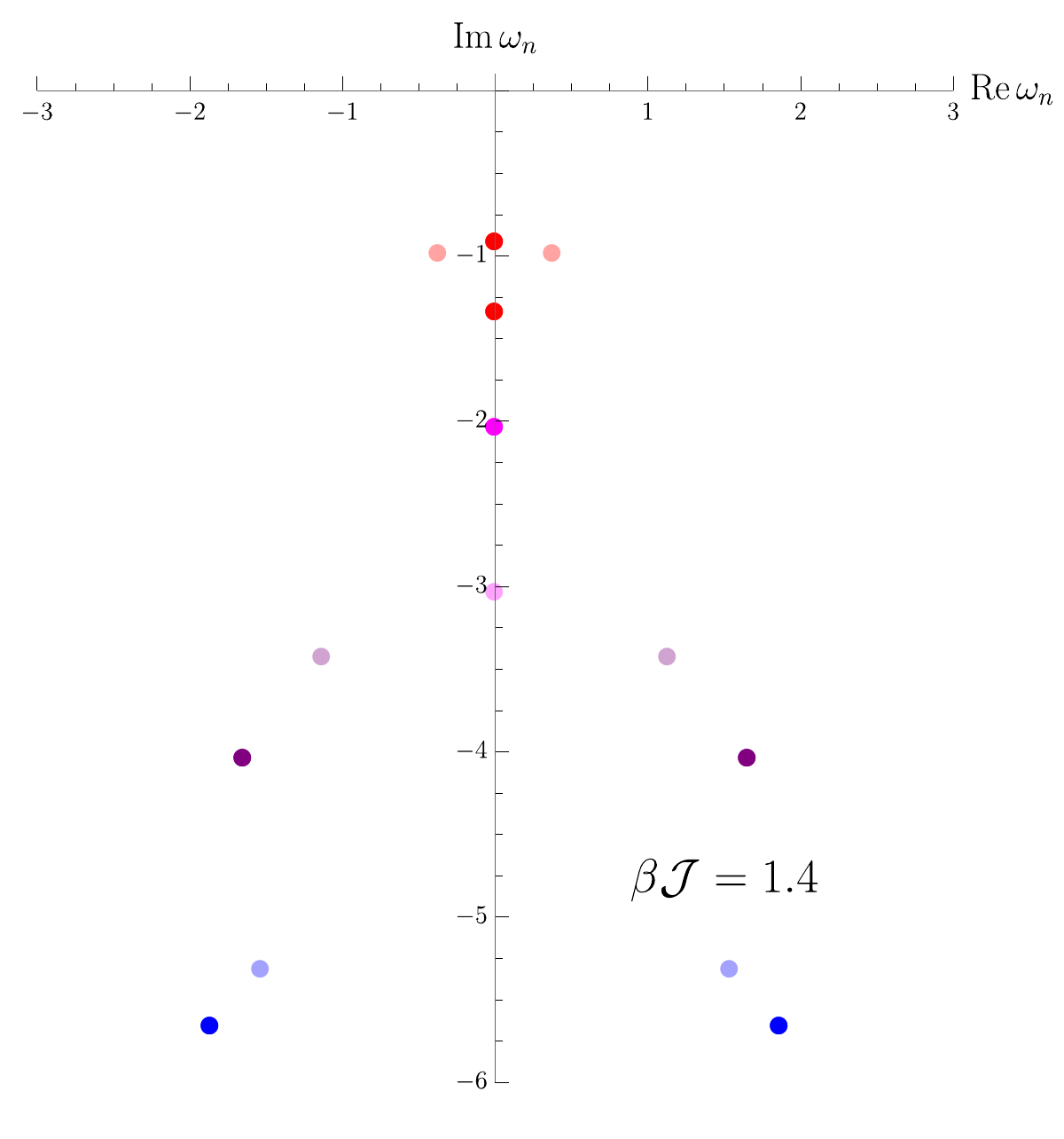}
    \includegraphics[width=0.32\linewidth]{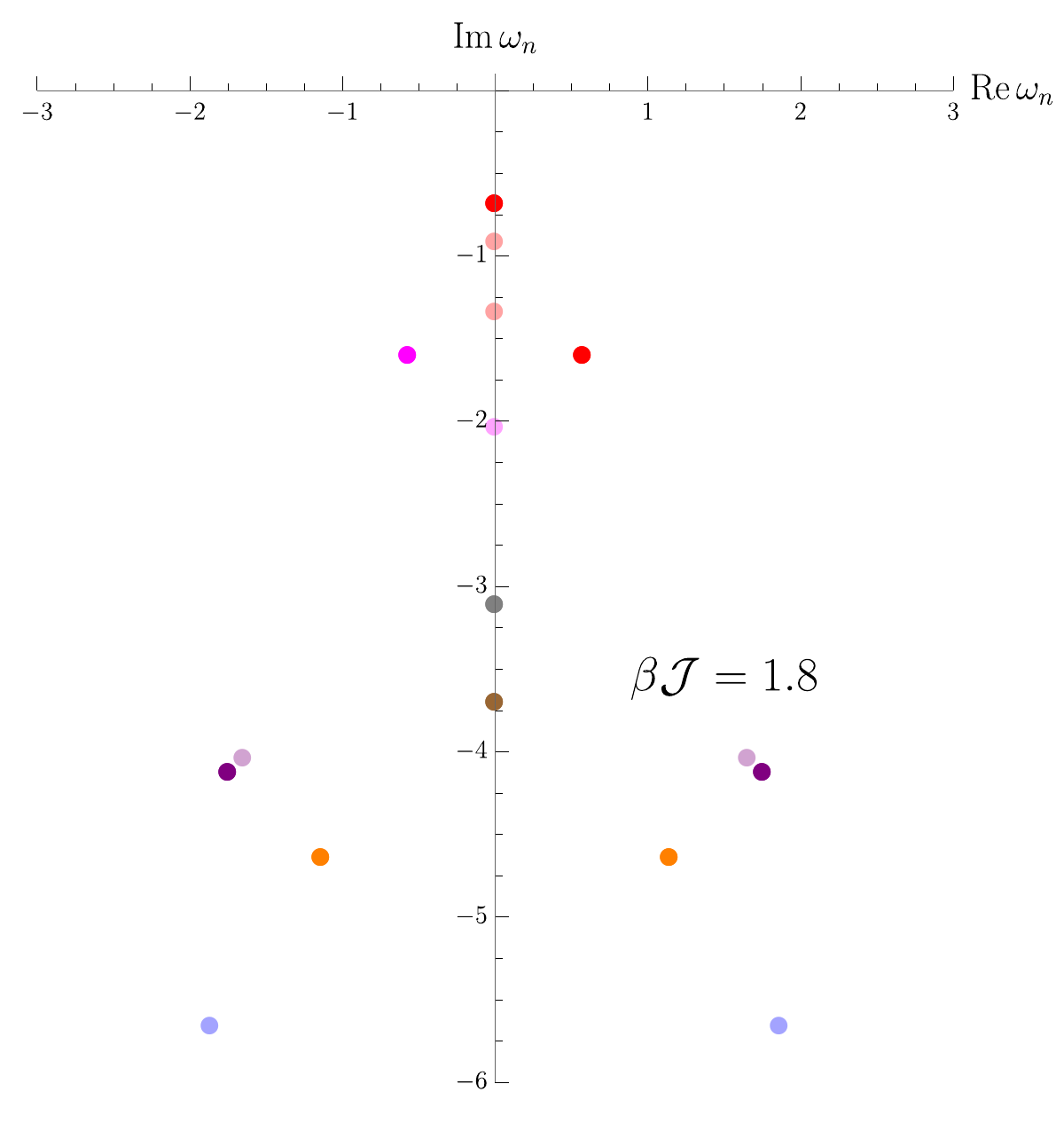}
    \includegraphics[width=0.32\linewidth]{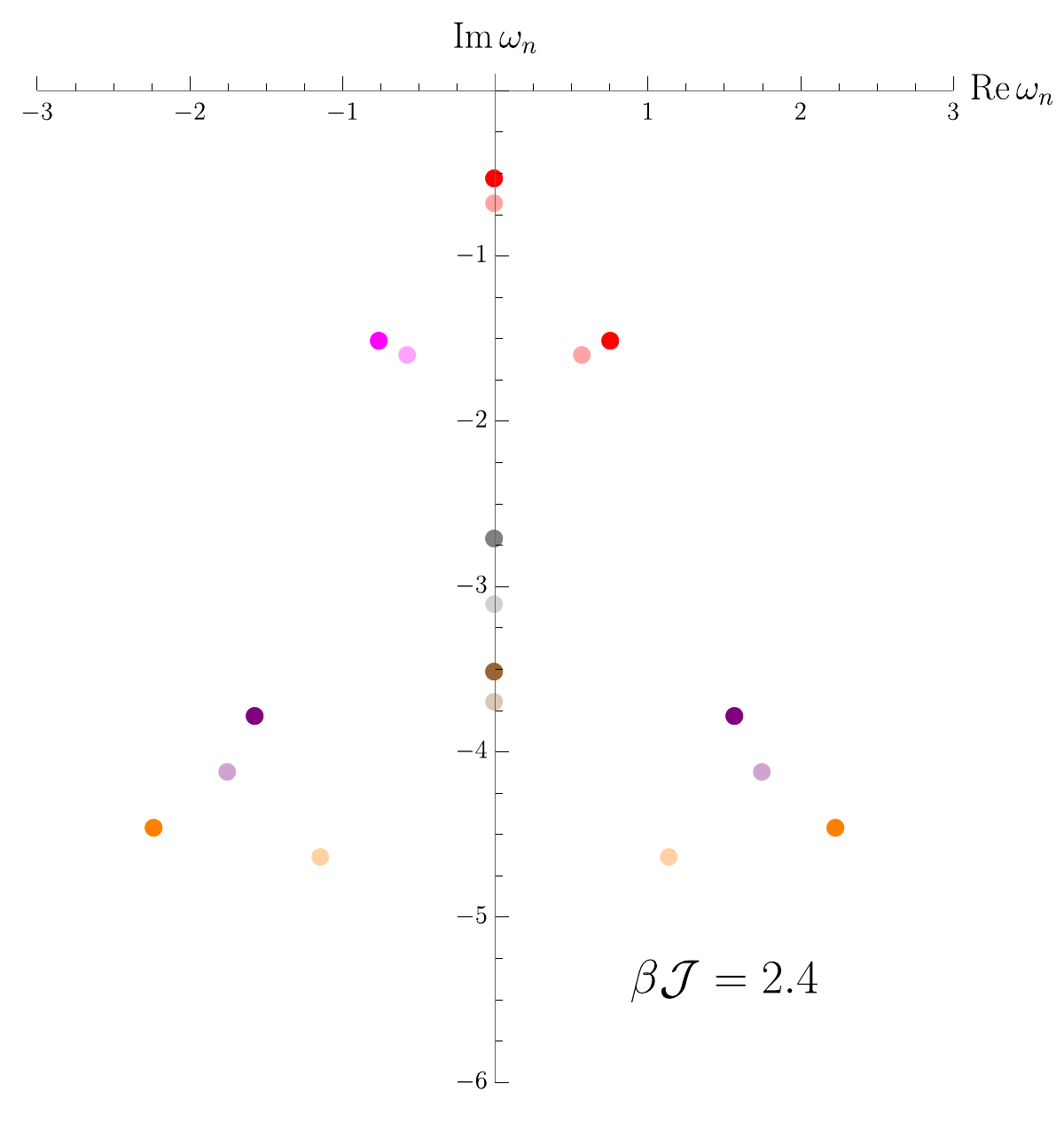}
    \includegraphics[width=0.32\linewidth]{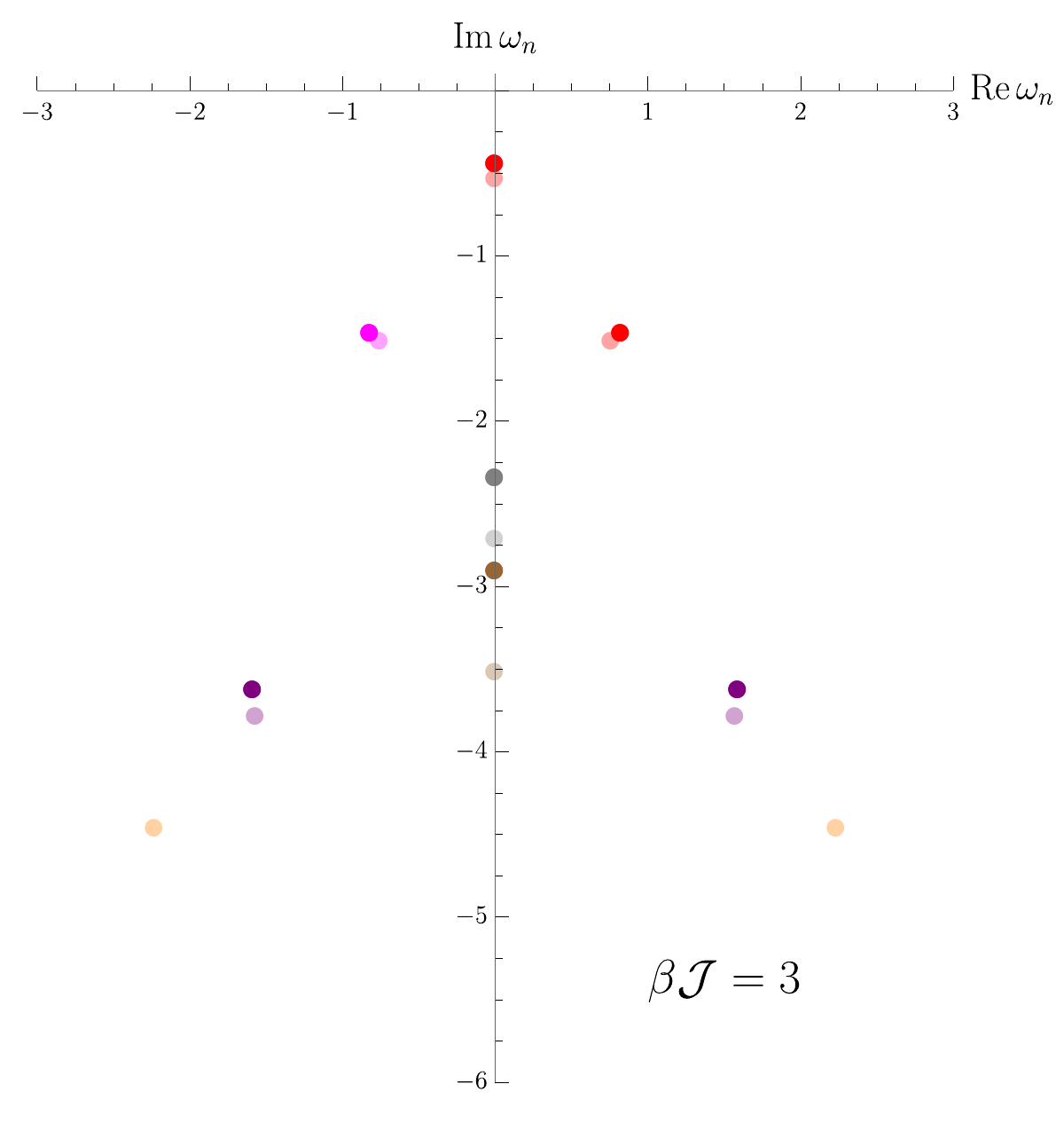}
    \includegraphics[width=0.32\linewidth]{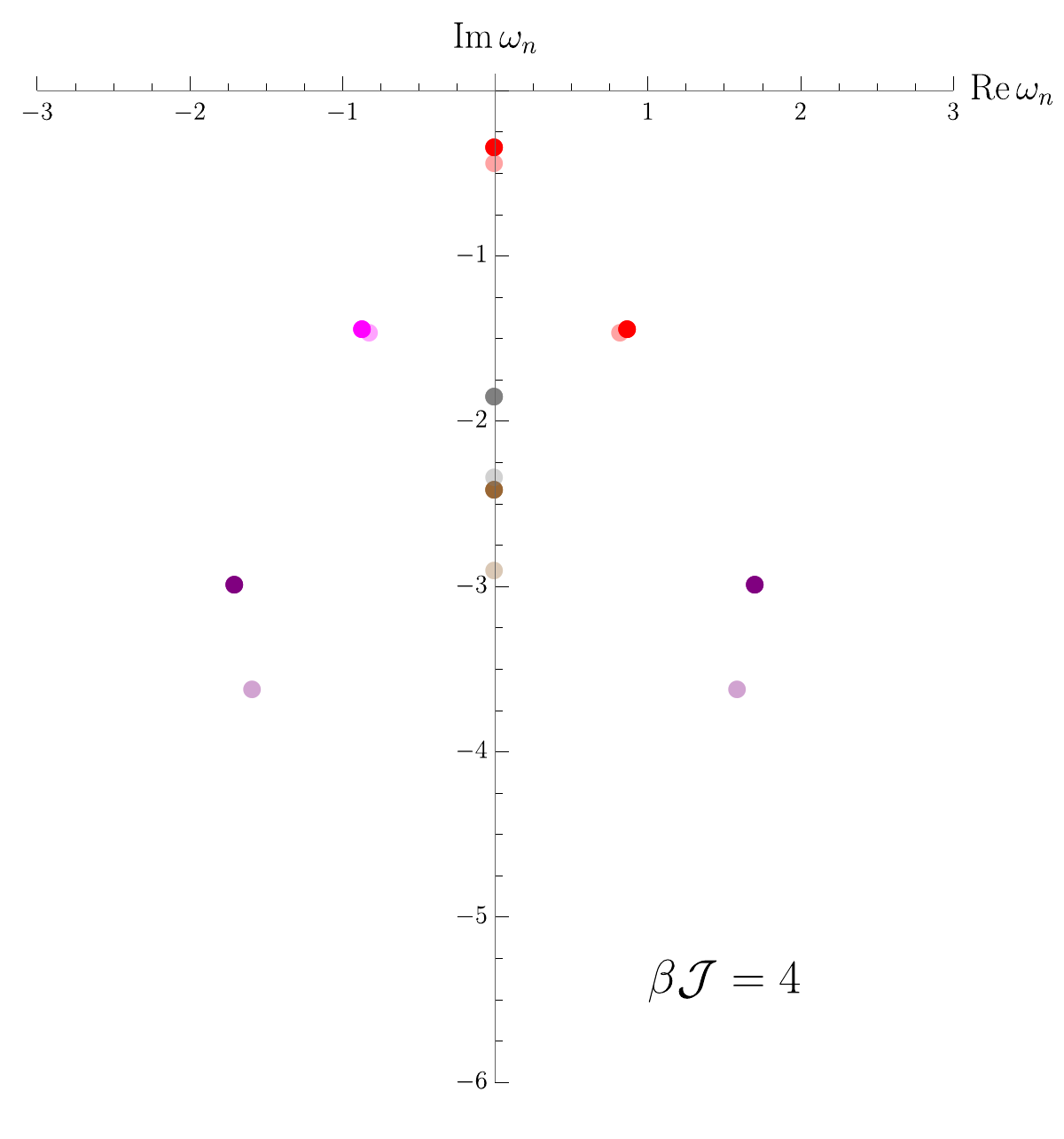}
    \includegraphics[width=0.32\linewidth]{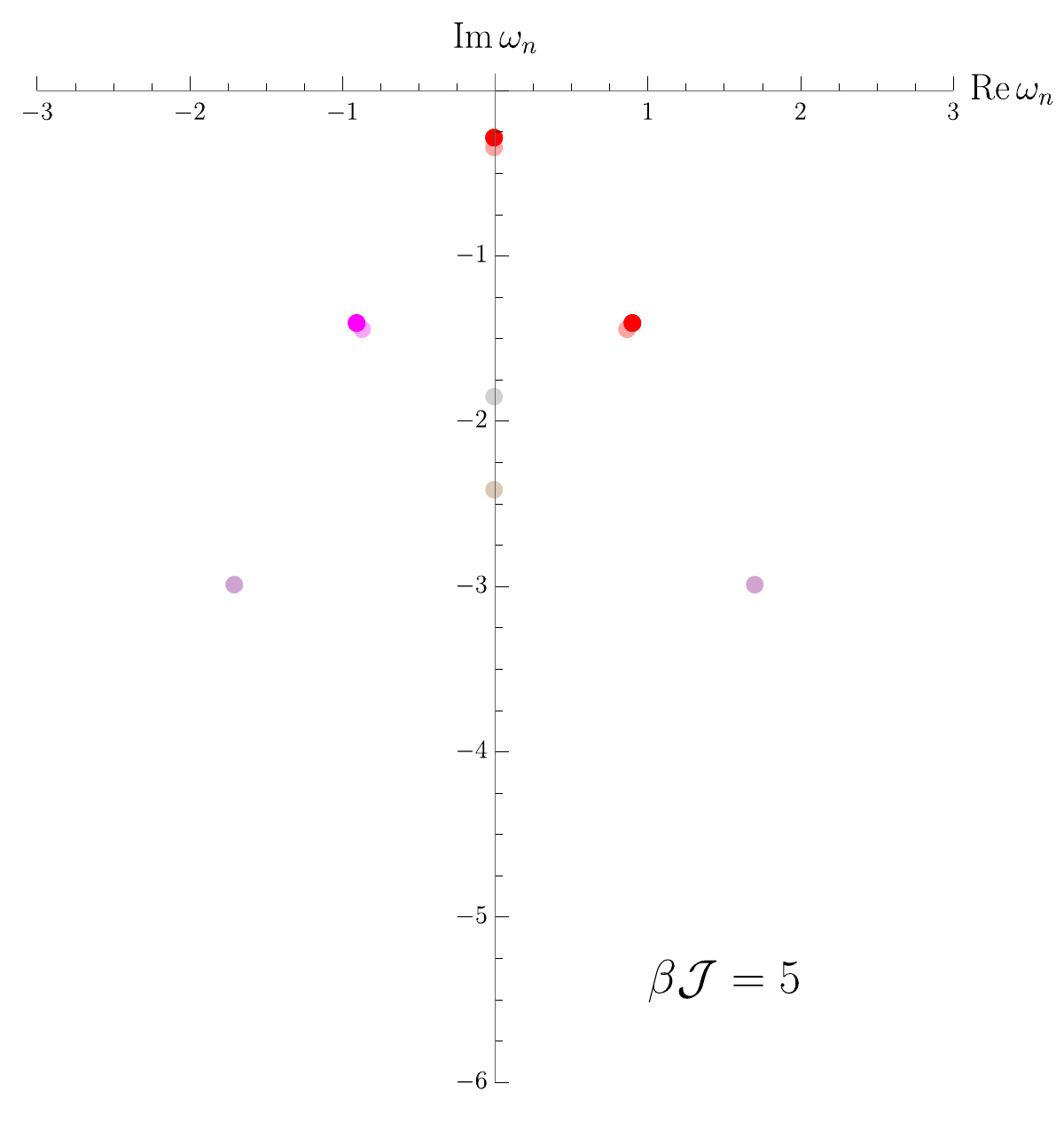}
    \caption{QNMs of the $p=4$ SYK model as a function of temperature. Each (pair of) QNM(s) that has converged in our scheme at a given value of $\betaJ$ is denoted using a solid circle of a given colour while faded circles of the same colour denote values from the previous panel. The panel with $\betaJ =0 $ is the same as Figure 3 from~\cite{Dodelson:2024atp} where the open circles denote QNMs that have not yet converged in our scheme. Initially, the QNMs trace out complicated trajectories but as one approaches the low temperature gravity regime, all QNMs begin to monotonically approach the real frequency axis in agreement with~\eqref{hotfast}.}
    \label{fig:betascan}
\end{figure}

Let us now consider the $p = 4$ spectrum. In Figure~\ref{fig:betascan}, we plot QNMs as a function of temperature in the range $0 \leq \betaJ\leq 5$. The first panel is identical to the infinite temperature result of Figure 3 in~\cite{Dodelson:2024atp} where the solid circles represent QNMs that have converged within our scheme, and the open circles represent QNMs that have not, but nevertheless we know are present from Figure 3 in~\cite{Dodelson:2024atp}. 

Close to $\betaJ=0$, we find the movement is anti-monotonic for \emph{all} QNMs, i.e.,~all QNMs move \emph{away} from the real frequency axis as the temperature is \emph{decreased}. We note also that the Christmas tree becomes narrower as temperature is decreased close to infinite temperature. 

For $\betaJ \sim 1$, the movement is more complicated. We note there are collisions between poles occurring on the first and second levels, and it is not possible to classify the motion of all QNMs as monotonic or anti-monotonic. 

\begin{figure}
    \centering
    \includegraphics[width=0.8\linewidth]{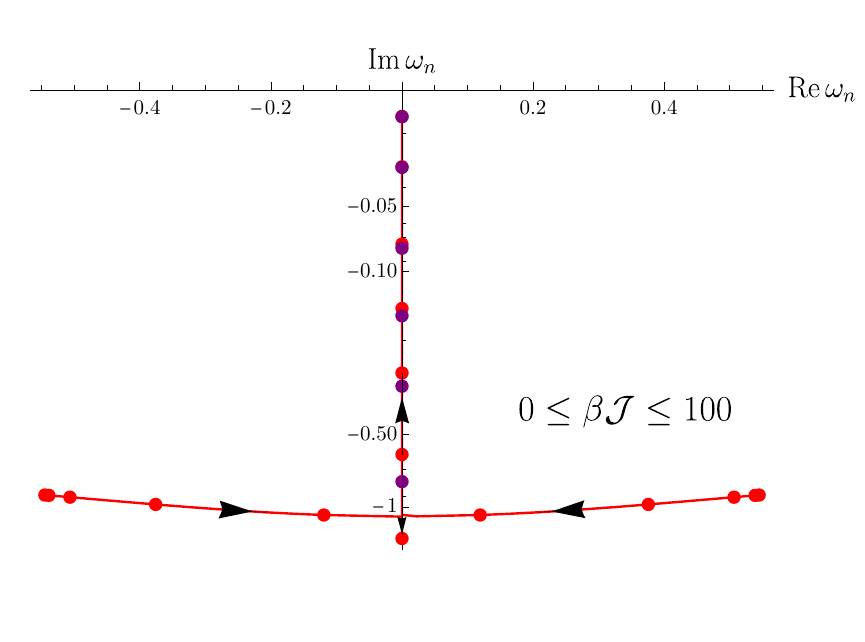}
    \includegraphics[width=0.8\linewidth]{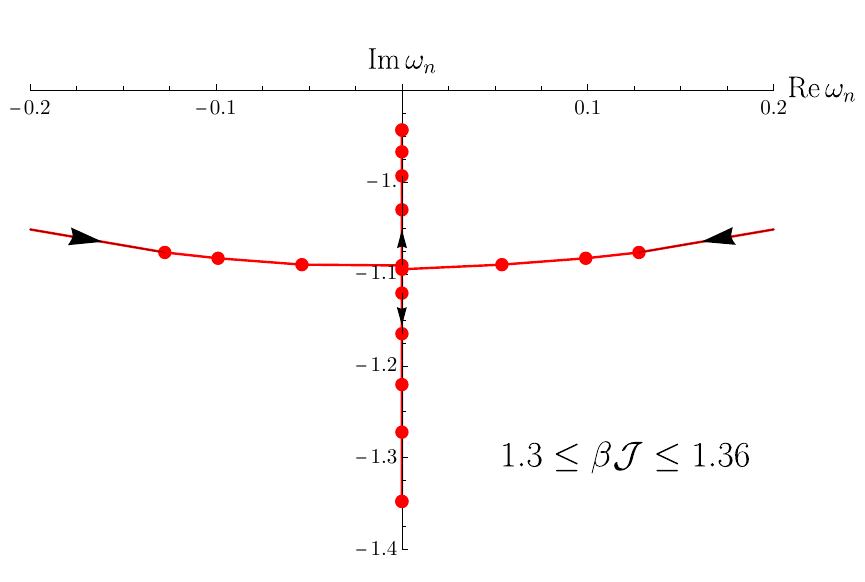}
    \caption{Trajectories of the leading pair of QNMs of the $p = 4$ SYK model (red) and the leading JT gravity QNM~\eqref{eq:JTQNMs} (purple) as a function of temperature. As discussed in the main text, we are able to track the leading QNM to large values of $\betaJ$. As a result, we can make a precise match with the leading JT gravity QNM. {\bf Top:} We highlight $\beta \mathcal{J} \in \{y\times 10^{j}:y\in\{0,2,5,10\},j\in\{-1,0,1\}\} \cup \{1.3\}$ while the JT gravity QNM is plotted for $\beta \mathcal{J} \geq 2$. {\bf Bottom:} We zoom in to the interval $\beta \mathcal{J} \in [1.3,1.36]$ where the poles collide with each other on the imaginary axis.}
    \label{fig:QNMflow}
\end{figure}

However, beyond $\betaJ \sim 1$, we note the movement becomes regular again --- as we approach the low temperature regime dual to JT gravity, \emph{all} QNMs are monotonic in accordance with~\eqref{hotfast}. Although there is still a Christmas tree visible at the temperatures we investigate, we note the QNMs on the imaginary axis (indicated in grey and brown) move up the imaginary axis quicker than QNMs on the complex plane. Next, our analysis of Figure~\ref{fig:QNMflow} will lend further credence to the idea that this spectrum should flow to the purely imaginary spectrum of JT gravity at very low temperatures. 

In Figure~\ref{fig:QNMflow}, we track, much more closely, the leading pair of QNMs in the range $0 \leq \betaJ\leq 100$. Initially, the pair describes an anti-monotonic trajectory and collides on the imaginary axis as seen in Figure~\ref{fig:betascan}. After collision, one of the poles goes on to further collide with the pole indicated in magenta in Figure~\ref{fig:betascan}, and returns to the complex plane. The other pole continues up the imaginary axis monotonically, and plotting the leading JT gravity QNM~\eqref{eq:JTQNMs} in purple, we see that there is an excellent match between the two at large $\betaJ$. Although we cannot show this precisely, our numerics also suggest that the QNM indicated in grey in Figure~\ref{fig:betascan} flows over to the second JT gravity QNM. 

Lastly, we note that in the normalisation we use, the JT gravity QNMs go as $\sim 1/\beta$ whereas the QNMs on the complex plane in Figure~\ref{fig:QNMflow} appear to stay at an approximately fixed $O(1)$ value beyond $\betaJ =1$. If this behaviour continues to hold into the very low temperature regime, in a normalisation where we multiply $\omega$ by $\beta$, we would expect the QNMs on the complex plane to move towards negative infinity and disappear from the spectrum leaving only the  equispaced QNMs on the imaginary axis.

Hence using QNMs, we have managed to interpolate, at finite $p$, between the stringy infinite temperature regime, and the holographic low temperature regime dual to JT gravity coupled to a massive fermion. 

\section{QNMs in other systems} \label{sec:examples}

In the previous section we have tracked the movement of QNMs in the $p=4$ SYK model as temperature is changed: the movement was complicated, but became regular in the gravity regime, where it obeys~\eqref{hotfast}. In this section we demonstrate in a variety of setups that QNMs of AdS black holes in many analytically and numerically tractable regimes obey the same inequality. 

However, in Sections~\ref{sec:hydropole} and~\ref{sec:SSB} we find two classes of examples connected to symmetries, where a QNM violates the inequality~\eqref{hotfast}. Since symmetries can be broken by an arbitrary small amount, and since an infinitesimal breaking would not reverse the movement of QNMs but would make them technically generic, we conclude that~\eqref{hotfast} cannot hold in full generality even in the gravity regime and even for just generic modes. 

We end the section with an analysis of QNMs in the energy density two-point function in the large $p$ SYK chain. In the low temperature limit, non-hydrodynamic QNMs obey the inequality~\eqref{hotfast}, but not the diffusion pole which has anti-monotonic movement.

\subsection{AdS black branes}
Our first example is a holographic theory on $\mathbf{R}^{d-1}\times \mathbf{S}^1_\beta$. The dual geometry is an AdS${}_{d+1}$ black brane with the Lorentzian metric given by
\begin{align}\label{branemetric}
ds^2=-f(r)\, dt^2+\frac{dr^2}{f(r)}+r^2\, d\vec{x}^2\,,\hspace{10 mm}f(r)=r^2-\frac{\mu}{r^{d-2}}\,,
\end{align}
where $\vec{x}=(x_1,\dots,x_{d-1})$ are coordinates in the directions along the brane. Here we have set the AdS radius to one. The temperature is related to the mass density $\mu$ via 
\begin{align}\label{tempfinvol}
T=\frac{d\mu^{1/d}}{4\pi}\,.
\end{align}
\indent For simplicity, we consider the case of a free scalar field with mass dual to the boundary conformal dimension $\Delta$ in the background~\eqref{branemetric}. The quasinormal modes are functions $\omega_n(k)$ of the spatial momentum $k$. For any fixed $n$ and $k$, the analogue of the monotonicity inequality \eqref{hotfast} in this case is
\begin{align}\label{monmom}
\frac{{d}|\text{Im}\,\omega_n(k)|}{{d}T}> 0\,.
\end{align} 
Dimensional analysis requires that $\omega_n(k)/T$ is a function of the dimensionless parameter $k/T$.\\
\indent Let us start by analysing several limits in which the quasinormal modes are analytically tractable. The first case is large mode numbers $n$, for which one finds a family of evenly spaced modes,
\begin{align}\label{largetlargen}
\omega_n=4\pi T\sin\left(\frac{\pi}{d}\right)e^{-\frac{i\pi}{d}}n+\mathcal{O}(1)\,.
\end{align}
Note that $\omega_n$ is independent of $k$ in this limit, so that the scaling $\omega_n\sim T$ is a consequence of dimensional analysis. The monotonicity property~\eqref{monmom} is then manifest.\\
\indent Another regime in which analytic expressions are known is at large momentum $k$. In this limit one finds the Bohr-Sommerfeld quantization condition 
\es{BSquant}{
\sqrt{k^2-\omega_n^2}\int_{r_T}^{\infty}dr\,\frac{\sqrt{1-\left(\frac{r_T}{r}\right)^d}}{r^2-\mu r^{2-d}}=\frac{\pi\left(\Delta-\frac{d}{2}+1+2n\right)}{2}\,,\hspace{10 mm}n=0,1,\ldots\,,
}
where the turning point is at $r_T^d=\mu/(1-(\omega_n/k)^2)$. For ${k/\mu^{1/d}}\gg \abs{n}\gg 1$ the QNMs were computed in~\cite{Festuccia:2008zx,Dodelson:2023nnr}:
\es{LargekExp}{
\om_n(k)=k-{i\mu^{2\ov d+2}\ov (-ik)^{d-2\ov d+2}}\le(\sqrt\pi \,\Ga\le(\frac32+{1\ov d}\ri)\, n\ov 2^{\frac12+{1\ov d}}\Ga\le(1+{1\ov d}\ri)\ri)^{2d\ov d+2}+\dots\,.
}
Since the imaginary part of this expression is $\text{Im }\omega_n(k)\propto -\mu^{2\ov d+2} \propto -T^{2d\ov d+2}$, monotonicity again follows. Additionally, in Appendix~\ref{app:BSproof} we provide a somewhat technical argument to prove the monotonicity of the imaginary part of $\omega_n(k)$ for any $n$ in the large $k$ regime starting from~\eqref{BSquant}.

\subsection{AdS black holes}

\indent Instead of a holographic theory on $\mathbf{R}^{d-1}\times \mathbf{S}^1_\beta$, now consider a theory on $\mathbf{S}^{d-1}\times \mathbf{S}^1_\beta$. The dual geometry is now an AdS black hole with the Lorentzian geometry given by
\begin{align}\label{branemetricfinvol}
ds^2=-f(r)\, dt^2+\frac{dr^2}{f(r)}+r^2\, d\Omega^2_{d-1}\,,\hspace{10 mm}f(r)=r^2+1-\frac{\mu}{r^{d-2}}\,,
\end{align}
where $d\Omega^2_{d-1}$ is the metric of the unit $(d-1)$-sphere. It dominates the canonical ensemble for $\mu>2$, and in this regime, the temperature $T=f'(r_0(\mu))/(4\pi)$ (where $r_0$ is the horizon radius) is a monotonically increasing function of $\mu$. Monotonicity therefore means
\begin{align}\label{monspin}
\frac{d|\text{Im }\omega_n(\ell)|}{d\mu}>0\,,\hspace{10 mm}\mu>2\,,
\end{align}
where $\ell$ is the spin. 
\\
\indent In this section, we analyse several limits in which the quasinormal modes are analytically tractable before giving some numerical results. The first case is large mode numbers $n$, for which one finds a family of evenly spaced modes \cite{Natario:2004jd,Cardoso:2004up},
\begin{align}\label{largen}
\omega_n=\frac{\pi}{z_0}n+\mathcal{O}(1)\,,
\end{align}
To define the prefactor $z_0$, let $r_n$ be the complex solutions to the equation $f(r)=0$. Then 
\begin{align}\label{zzero}
z_0=\sum_{n=0}^{d-1}\frac{1}{f'(r_n)}\log\left(-\frac{1}{r_n}+i\epsilon \right)\,,
\end{align}
where $r_0>0$ is the black hole horizon radius.
In the large $\mu$, high temperature limit we recover the black brane physics from the previous section.~\eqref{largen} reduces to~\eqref{largetlargen} and monotonicity follows.

\begin{figure}[t]
\centering
\includegraphics[scale=.8]{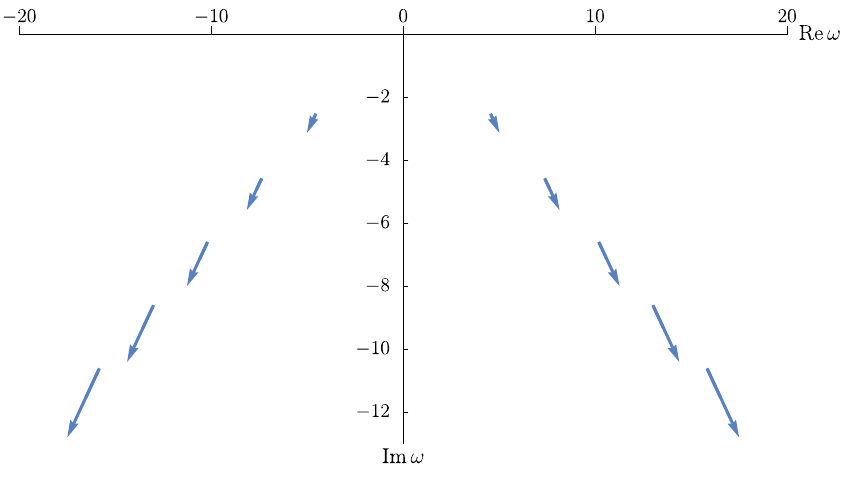}
\caption{The trajectories of the first five quasinormal modes in $d=4$ for $\ell=0$ and $\Delta=4$ as $\mu$ increases from $2$ to $3.5$. The arrows are in the direction of increasing $\mu$.\label{finvolmodes}}
\end{figure}

At smaller $\mu$, using~\eqref{zzero} one can show in $d=4$ that
\es{4dres}{
{\rm Im}\le({1\over z_0}\ri)=-{2r_0(\mu)\ov \pi}\,,\qquad \text{(for $d=4$)}
}
which immediately gives monotonicity. In other dimensions
it is straightforward to check numerically that $|\text{Im }\omega_n|$ monotonically increases from 0 at $\mu=0$ to~\eqref{largetlargen} at high temperatures, thereby confirming the expectation~\eqref{monspin}.
\\
\indent Another regime in which analytic expressions are known is at large spin $\ell$. In this limit, the quasinormal modes become long-lived quasiparticles, and the imaginary part of the frequencies is related to the probability to tunnel over the centrifugal barrier \cite{Festuccia:2008zx,Dodelson:2022eiz}, 
\begin{align}
|\text{Im }\omega_n(\ell)|\propto\exp\left(-2\int_{r_-}^{r_+}\frac{dr}{f(r)}\, \sqrt{\frac{\ell^2 f(r)}{r^2}-(\text{Re }\omega_n(\ell))^2}\right)\,,
\end{align}
where $r_{\pm}$ are the two turning points outside the horizon. The real part of the frequency satisfies the constraint
\begin{align}\label{realconstraint}
\ell<|\text{Re }\omega_n (\ell)|<\Omega\ell\,,
\end{align}
where $\Omega$ is the velocity of null geodesics at the photon sphere, 
\begin{align}
\Omega=\sqrt{1+\left(1-\frac{2}{d}\right)\left(\frac{2}{d\mu}\right)^{\frac{2}{d-2}}}\,.
\end{align}
Differentiating with respect to $\mu$, we find
\begin{align}
\frac{d\log |\text{Im }\omega_n(\ell)|}{d\mu}=\int_{r_-}^{r_+}\frac{dr}{r^{d-2}f(r)^2}\frac{2(\text{Re }\omega_n(\ell))^2-\frac{\ell^2f(r)}{r^2}}{\sqrt{\frac{\ell^2 f(r)}{r^2}-(\text{Re }\omega_n(\ell))^2}}\,.
\end{align}
For $\omega_n$ in the range~\eqref{realconstraint} and $\mu>2$, it is straightforward to check that the integrand is manifestly positive. Therefore monotonicity is again satisfied.\\
\indent For generic values of the spin and mode number, the quasinormal modes must be computed numerically. In Figure~\ref{finvolmodes}, we display the spectrum of modes in $d=4$ for $\ell=0$ as a function of $\mu$, computed using the Mathematica package $\mathtt{QNMSpectral}$~\cite{Jansen:2017oag}. The modes are monotonic in temperature for all parameter choices we have checked, so the inequality~\eqref{hotfast} is again respected. Note that modes further down in the complex frequency plane vary more strongly as a function of temperature. This is related to the quasinormal instability discussed in~\cite{Boyanov:2023qqf,Cownden:2023dam,Arean:2023ejh}.

\subsection{Charged black branes}

Let us consider another example: the charged black brane in AdS$_{5}$. Its metric is given by
\begin{align}
ds^2=-f(r)\, dt^2+\frac{dr^2}{f(r)}+r^2\, d\vec{x}^2\,,\hspace{10 mm}f(r) ={r^2}-\frac{\mu}{r^{2}}+\frac{Q^2}{r^{4}}\,.
\end{align}
The horizon is located at
\begin{align}
    f(r_0)=0\implies r_0 = r_0(\mu,Q)\,,
\end{align}
where $r_0$ is a complicated function of $\mu$ and $Q$ that can be solved for in closed form. From this, we can obtain the expression for the temperature in terms of $\mu$ and $Q$ as
\begin{align}
    T = \frac{1}{4\pi}\left.\frac{df(r)}{dr}\right |_{r=r_0(\mu,Q)}\,.
\end{align}

\begin{figure}[!h]
\centering
\includegraphics[scale=0.7]{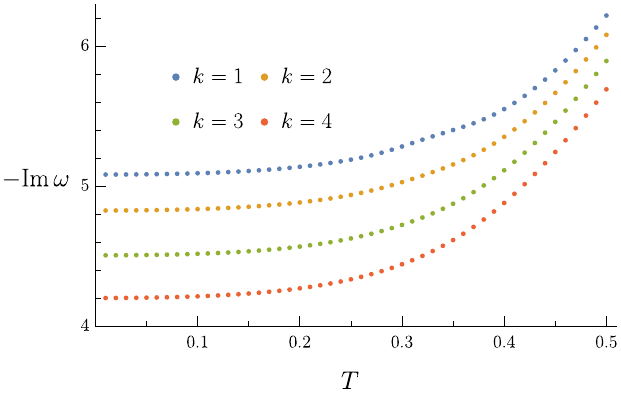}\hspace{0.5cm}
\includegraphics[scale=0.7]{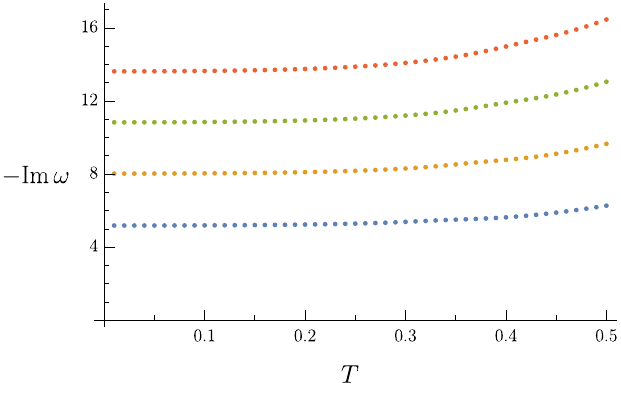}
\caption{The temperature dependence of the imaginary part of the first few QNMs  in the charged AdS$_5$ black brane for a scalar field with $\Delta=5$, at fixed $Q=1$. {\bf Left:} We plot the first QNM (the one with the smallest $|\text{Im }\omega|$) at different values of $k$. {\bf Right:} We plot the first four QNMs at $k=0$.\label{fig:RN5D}}
\end{figure}

We again consider a scalar field with conformal dimension $\Delta$. For fixed $\Delta$ and $Q$, we can study the temperature dependence of the QNMs by varying $\mu$. We plot the temperature dependence of a few example QNMs in Figure~\ref{fig:RN5D} and see that they all obey the inequality~\eqref{hotfast}.

\subsection{Hydrodynamic poles} \label{sec:hydropole}

We expect generic quantum systems to have slow modes only if there is a symmetry reason for their existence. The transport of conserved densities proceeds through hydrodynamic modes that become slow at long wavelengths. If the system breaks a continuous symmetry spontaneously, it will have Goldstone bosons. Below, we will consider two examples, where both kinds of modes lead to the violation of monotonicity~\eqref{hotfast}. Note that we are not arguing for the converse: slow modes can obey monotonicity.

One may try to refine the conjecture of monotonicity by excluding symmetry protected soft modes. In general, this would not work, as breaking a symmetry very weakly would not alter the monotonicity-violating dispersion relation of such modes significantly, but would make these modes technically not symmetry protected. 

In~\eqref{DiffDisp} we provided a dimensional analysis argument for the diffusion constant, and hence the purely imaginary QNM scaling as $D={c_D/  T}$, which violates monotonicity. Here we present an intuitive argument for this scaling due to Luca Delacr\'etaz.

Let us consider a charge density long wavelength wave with associated wave number $k$.  Because the charge is conserved, this configuration can only relax by the transport of charge from the high density region to the low density region. Assume that quasiparticles transport the charge and $T\gg m$ so that they are relativistic: their mean free path and mean free time scales as $\lam_\text{MFP}\sim 1/T$, they perform a random walk, hence in time $t$ they scatter $n=t/\lam_\text{MFP}$ times, they will have travelled an average distance $d$ obeying:
\es{dav}{
d^2=n \lam_\text{MFP}^2=t \lam_\text{MFP}\,.
}
The distance $d$ has to match $1/k$, and $t\sim 1/\abs{\om}$ (with the absolute value needed because the charge does not oscillate, but decays), hence we obtain 
\es{dav2}{
\abs{\om} \sim \lam_\text{MFP} k^2\sim {k^2\ov T} \,.
}
Thus we obtain the scaling of the diffusion constant $D\sim 1/T$. Note that in a weakly-coupled QFT there would be important factors of the coupling constant in the above computation, but this would not modify the scaling with temperature.

\subsection{An example with symmetry breaking}
\label{sec:SSB}
\begin{figure}
    \centering
    \includegraphics[width=0.46\linewidth]{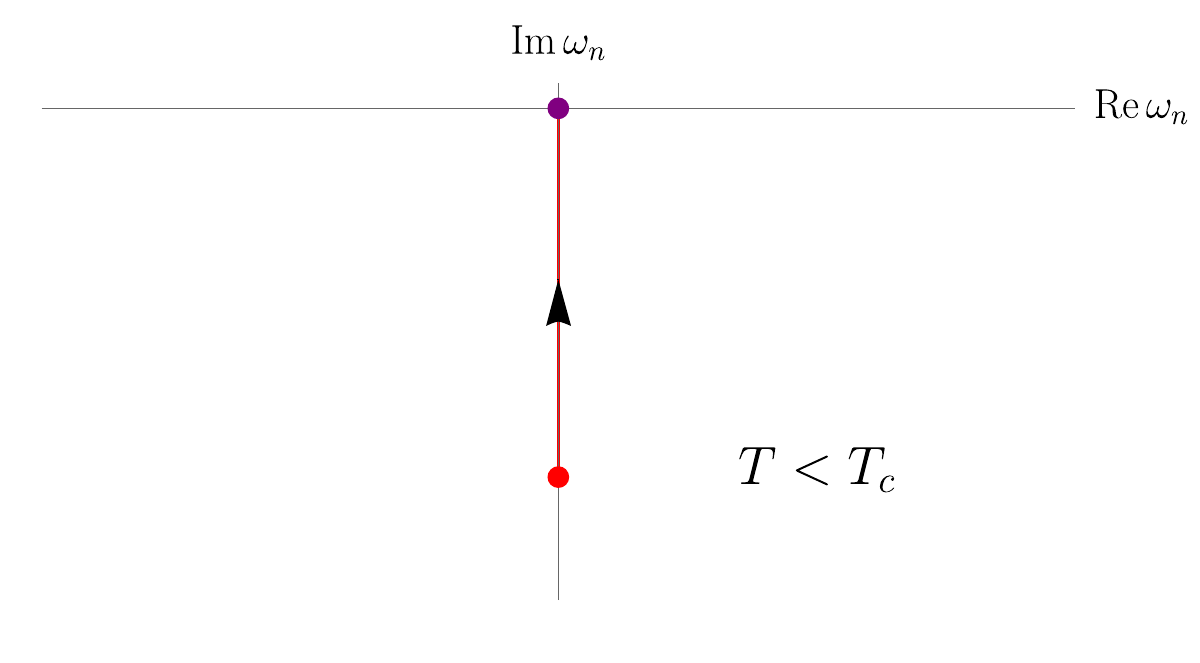}
    \includegraphics[width=0.46\linewidth]{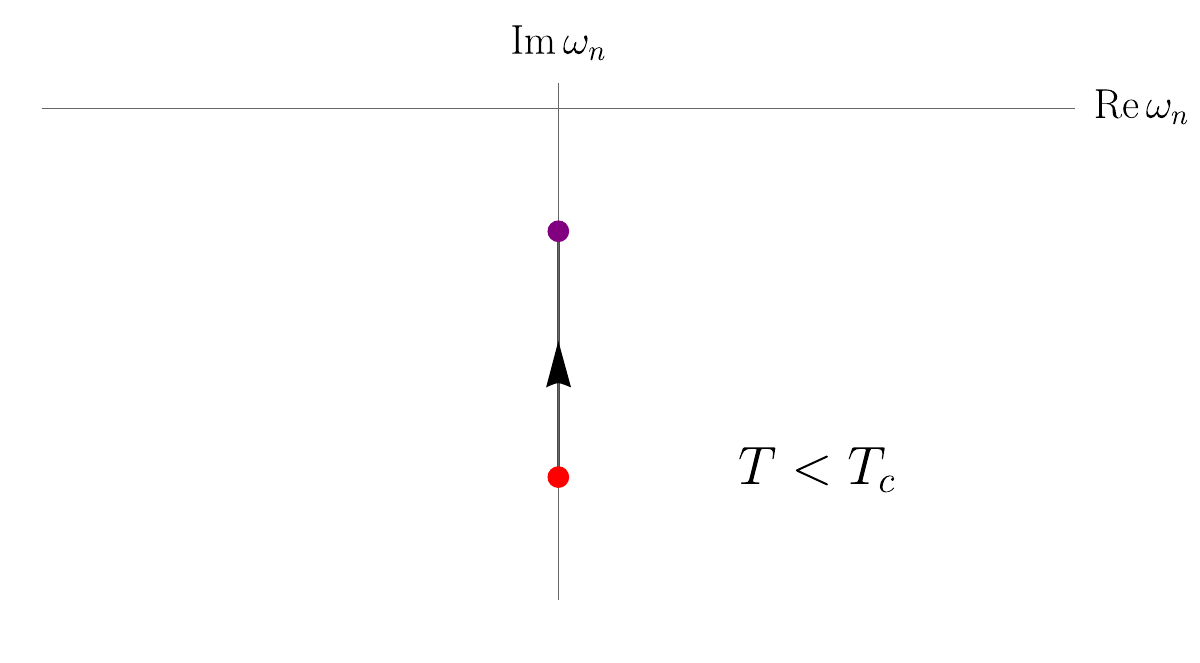}
    \includegraphics[width=0.46\linewidth]{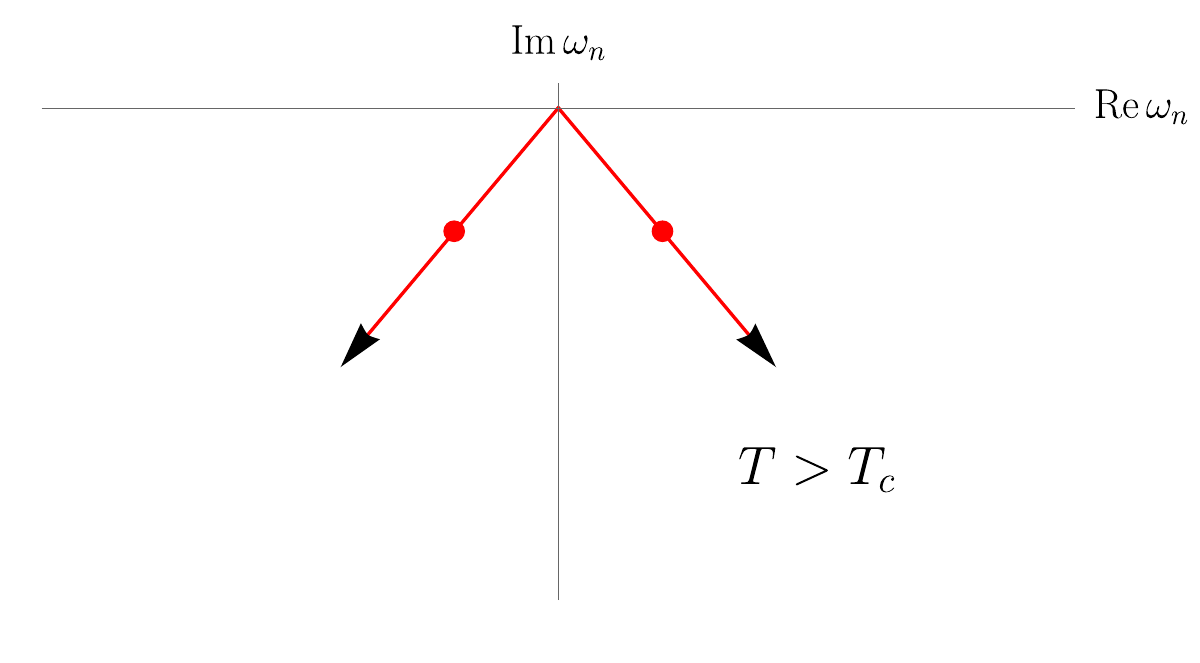}
    \includegraphics[width=0.46\linewidth]{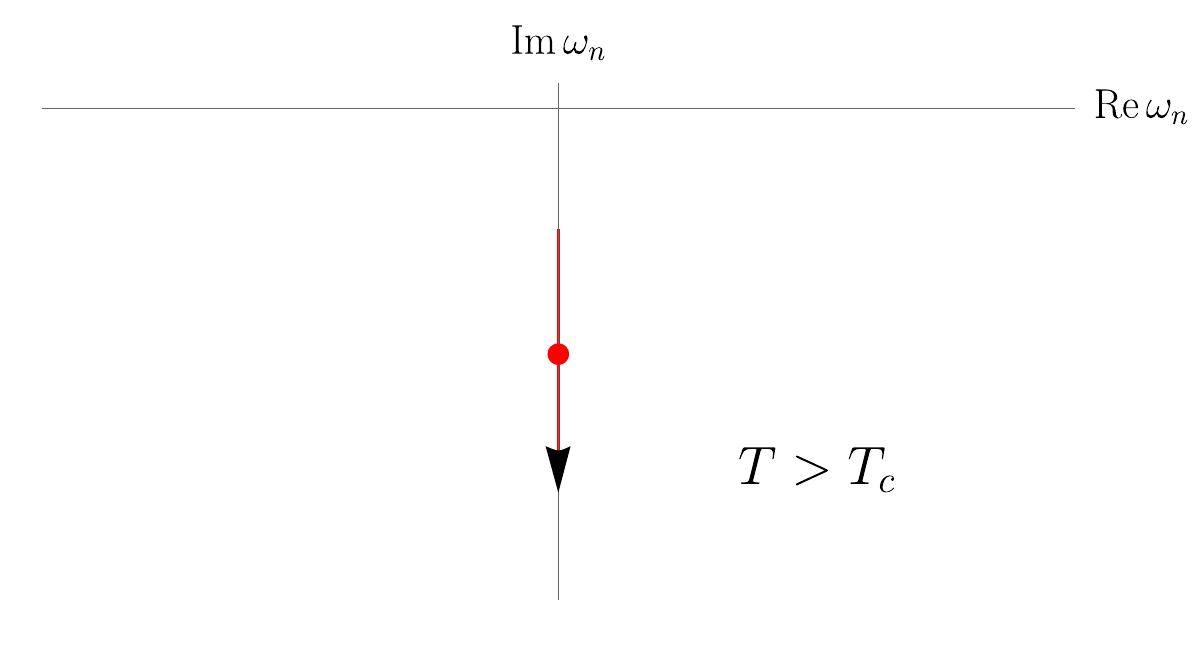}
    \caption{Cartoons of the trajectories in the complex plane of the leading pair of QNMs in the holographic superfluid setting of~\cite{Bhaseen:2012gg} (left) and the Model A universality class (right) as the temperature $T$ is varied. We consider trajectories linearised around the equilibrium state such that we zoom in around the Goldstone pole highlighted in purple. Note that the arrows point in the direction of increasing $T$ as opposed to Figure~\ref{fig:QNMflow} where they point in the direction of increasing $\beta$. As noted in the main text, the trajectory of the poles for $ T < T_c$ violates monotonicity~\eqref{hotfast}.}
    \label{fig:Goldstone}
\end{figure}
If a system breaks a continuous symmetry, the corresponding Goldstone boson is a gapless excitation, and  at $k=0$ gives a QNM sitting at the origin. When the symmetry is restored at $T=T_c$, this QNM has to be absent from the origin. In the setting where the Green's functions are meromorphic, the mechanism for this is that the Goldstone pole collides with another pole at $T=T_c$, and together they reorganise and move into the lower half plane. \emph{Since the colliding pole has to approach the origin as $T$ increases, it violates monotonicity~\eqref{hotfast}.}

We briefly discuss two examples. In the holographic superfluid setting in~\cite{Bhaseen:2012gg} it was found that after the collision the poles move off the imaginary axis; see the left plot on Figure~\ref{fig:Goldstone}.
The physics of this recombination can be understood without reference to holography: it is a dynamic critical phenomenon described by the universality class of Model F in the Hohenberg-Halperin classification~\cite{Hohenberg:1977ym}. Instead of going through this somewhat lengthy analysis, we consider the simplest Model A universality class, which also provides a counterexample to monotonicity.

Model A describes the complex $U(1)$ order parameter dynamics through the Langevin equation:
\es{ModelA}{
\p_t \phi(t,x)&= - \Ga\,{\de F\ov \de \phi(t,x)}+\xi(t,x)\,,\\
F[\phi]&=\int dx \,\le(\abs{\nabla\phi}^2+r\abs{\phi}^2+{u\ov 2}\abs{\phi}^4\ri)\,,\\
\langle \xi(t,x) \bar \xi(t',x') \rangle&=2\Ga\, T\, \de(x-x')\de(t-t')\,.
}
Near the phase transition $r=\al (T-T_c)+\dots$, while the temperature dependence of $u,\,\Ga$ is unimportant.
Note that Model F in addition to $\phi$ contains the conserved density leading to a coupled set of equations.

Linearising around the equilibrium state $\phi=\sqrt{\abs{r}\ov u}$ for $T<T_c$, we obtain the Goldstone and the amplitude modes:
\es{modes_belowTc}{
\om_\text{Goldstone}&=-i\Ga\, k^2\,,\\
\om_\text{amp}&=-i\Ga\, (2\abs{r}+k^2)\,.
}
Linearising around the equilibrium state $\phi=0$ for $T>T_c$, we obtain two relaxational modes:
\es{modes_belowTc2}{
\om_\text{rel}&=-i\Ga\, (r+k^2)\,.
}
 In contrast to the holographic example, Model A remains purely dissipative, so after the collision at $T=T_c$ (for which $r=0$) the poles stay on the imaginary axis. The movement of the poles is plotted on the two figures on the right on Figure~\ref{fig:Goldstone}.

\subsection{QNMs of the large $p$ SYK chain}
In the large $p$ SYK chain, analytical results are available for the two-point functions of fermions and energy density. The two-point function of fermions is ultra-local: it is equal to~\eqref{eq:largeqtwo} for the same site and it vanishes otherwise~\cite{Gu_2017}. However, a non-trivial example is provided by the retarded energy-energy correlator found in closed form in~\cite{Choi:2020tdj}. Restoring the temperature dependence in their results, we find
\begin{equation}
    G_R^{\epsilon \epsilon}(\omega, k) = \frac{- N v}{2q^2} \left.\left( \left.\partial_{\theta} \log \psi_n(\theta)\right|_{\theta = \theta_v} + \tan \frac{\pi v}{2} \frac{h (h-1)}{2} \right) \right|_{n \to - i \omega + \epsilon}\,, \label{eq:retenergy}
\end{equation}
where
\begin{align}
    \begin{split}
        \psi_n(\theta) & = c_o \psi_n^o(\theta) + c_e \psi_n^e(\theta) \,, \\
        c_o & = \frac{\Gamma \left( 1- \frac{h}{2} - \frac{n \beta}{4\pi v}\right) \sin \left(\frac{\pi h}{2}+ \frac{n \beta}{4\pi} \right) \sin \left(\frac{n \beta}{4}\right)}{\Gamma \left( \frac{1}{2} - \frac{h}{2}- \frac{n \beta}{4\pi v}\right)} \,,\\
        c_e & = \frac{\Gamma \left( \frac{1}{2}- \frac{h}{2} - \frac{n \beta}{4\pi v}\right) \cos \left(\frac{\pi h}{2}+ \frac{n \beta}{4\pi} \right) \cos \left(\frac{n \beta}{4}\right)}{2 \Gamma \left( 1 - \frac{h}{2}- \frac{n \beta}{4\pi v}\right)} \,,\\
        \psi^e_n (\theta) &= \sin^h \theta\, _2F_1 \left(\frac{h - n \beta / 2\pi v}{2}, \frac{h + n \beta / 2\pi v}{2}, \frac{1}{2}, \cos^2 \theta \right) \,,\\
        \psi^o_n (\theta) &= \cos \theta \sin^h \theta\, _2F_1 \left(\frac{1 + h - n \beta / 2\pi v}{2}, \frac{1 + h + n \beta / 2\pi v}{2}, \frac{3}{2}, \cos^2 \theta \right) \,,\\
        h(k) &= \frac{1}{2} \left( 1+ \sqrt{9 + 4 \gamma (\cos k - 1)} \right) \,,\qquad 
        \theta_v = \frac{\pi}{2}(1-v) \,,
    \end{split}
\end{align}
where $\gamma$ is the strength of the intersite coupling and $v$ was defined in~\eqref{eq:veq}.
\begin{figure}
    \centering
    \includegraphics[width=0.48\linewidth]{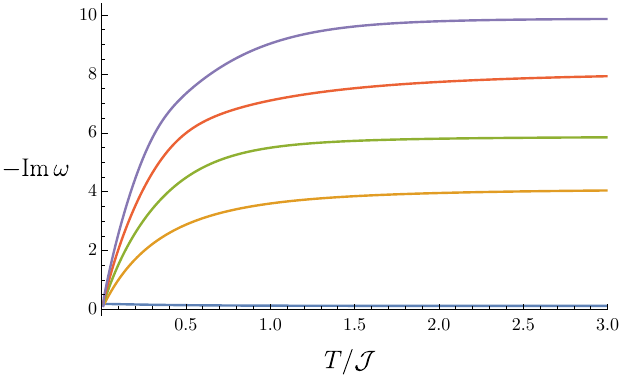}
    \includegraphics[width=0.48\linewidth]{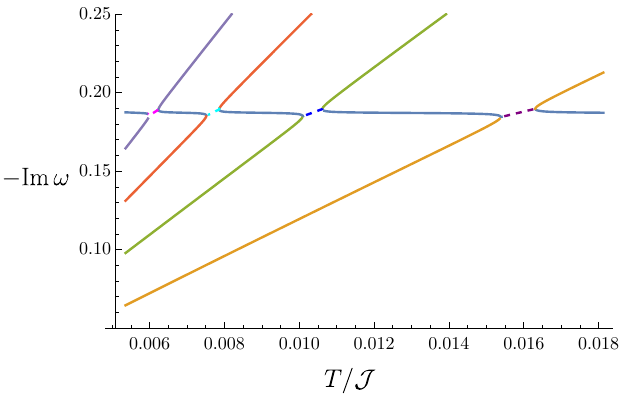}
    \caption{Plots of the five leading QNMs of the retarded energy density correlator~\eqref{eq:retenergy} $G^{\epsilon \epsilon}_R(\omega)$ for $k = 0.6$, $\gamma =1$. On the right, we zoom in around $T = 0$ where collision between the diffusion pole and a higher pole can occur on the imaginary axis (see Figure~8 in~\cite{Choi:2020tdj}). After collision, the poles trace a trajectory as complex conjugates in the complex plane (denoted using dashed lines) before recombining on the imaginary axis. We observe that the diffusion pole obeys a non-monotonic trajectory (the same way as in Section~\ref{sec:hydropole}) but that the non-diffusion poles obey monotonic trajectories in agreement with~\eqref{hotfast}.}
    \label{fig:chain-plots1}
\end{figure}
\begin{figure}
    \centering
    \includegraphics[width=0.48\linewidth]{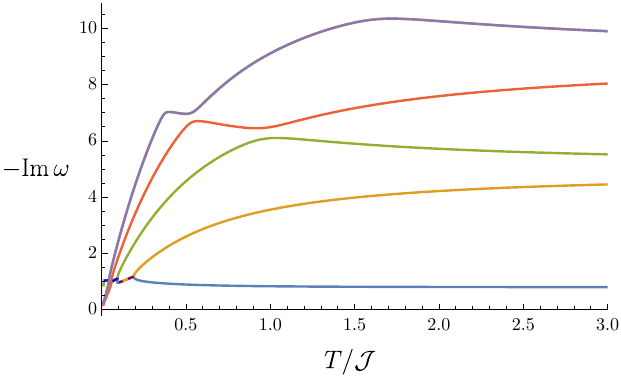}
    \includegraphics[width=0.48\linewidth]{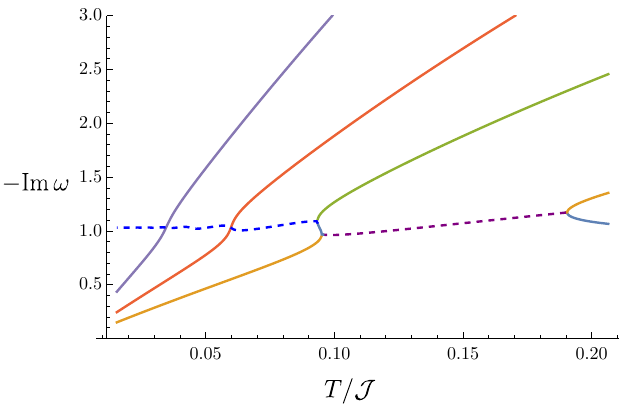}
    \caption{Plots of the five leading QNMs of the retarded energy density correlator~\eqref{eq:retenergy} $G^{\epsilon \epsilon}_R(\omega)$ for $k = 1.6$, $\gamma =1$. On the right, we zoom in around $T = 0$. In contrast to Figure~\ref{fig:chain-plots1} and reminiscent of Section~\ref{sec:spectrum}, the large/moderate $T$ trajectories are non-monotonic. Also in contrast to Figure~\ref{fig:chain-plots1}, the colliding trajectories are non-monotonic also. Nevertheless, the movement of QNMs becomes regular in the low $T$ holographic regime.}
    \label{fig:chain-plots}
\end{figure}

Then QNMs can be found numerically, as in Figures~\ref{fig:chain-plots1} and~\ref{fig:chain-plots}, as zeroes of the function $\psi_n(\theta)$ for some fixed $k, \gamma$, and as functions of $v$ or equivalently, the temperature $T$. The chain has a conserved energy density governed by a diffusion pole. While unlike in CFTs there is a dimensionful scale ${\cal J}$ in the problem and hence  the dimensional analysis argument around~\eqref{DiffDisp} loses power, we nevertheless find that the diffusion pole is anti-monotonic:
 the diffusion pole has an analytic expression for $\omega, k\ll 1$~\cite{Choi:2020tdj},
\es{SYKchain_diff}{
    \omega_D &\simeq -i D\,k^2\,,\\
    D&=\frac{\gamma\mathcal{J}}{6}\cos \frac{\pi v}{2}\left(\pi v \tan \frac{\pi v}{2} + 2 \right)\,.
} 
Then the temperature dependence is encapsulated in $D$, which is a monotonically increasing function of $v$. Since $v$ is a proxy for $\beta$ (with $v=0$ being infinite $T$ and $v\to 1$ low $T$), the diffusion pole is anti-monotonic for \emph{all} $T$. Indeed, this is what we find in Figures~\ref{fig:chain-plots1} and~\ref{fig:chain-plots} where the diffusion pole is plotted in blue. This behaviour is the same that we described in Section~\ref{sec:hydropole} for quasiparticle systems.

Other than the diffusion pole, Figures~\ref{fig:chain-plots1} and~\ref{fig:chain-plots} also plot the next four leading QNMs as a function of temperature for two momenta $k = 0.6,\,1.6$ at $\gamma = 1$. The QNMs indicated are all on the imaginary axis unless they collide with the diffusion pole after which they trace out a trajectory on the complex plane as complex conjugates (indicated using dashed lines) and recombine on the imaginary axis (see Figure~8 in~\cite{Choi:2020tdj}).

For Figure~\ref{fig:chain-plots1} with small $k$, i.e., $k = 0.6$, we note that the non-diffusion poles \emph{all} obey monotonic trajectories for \emph{all} temperatures in agreement with~\eqref{hotfast}. This includes the low $T$ regime where they collide with the diffusion pole, trace out a trajectory on the complex plane, recombine, and continue up the imaginary frequency axis. 

The story at $k = 1.6$ is more involved. At large/moderate $T$, the non-diffusion poles obey non-monotonic trajectories reminiscent of the movement of QNMs in Section~\ref{sec:spectrum}. Even when they collide with the diffusion pole, the trajectory is not necessarily monotonic --- see the dashed blue line in Figure~\ref{fig:chain-plots}. We note also that not all non-diffusion poles collide with the diffusion pole. Nevertheless, despite this complicated movement away from low $T$, the movement becomes monotonic as we approach the holographic regime in accordance with~\eqref{hotfast}.
\section{New results on operator growth}\label{sec:opgrowth}

Operator growth under time evolution is a sensitive probe of many-body quantum chaos. The two leading measures of operator growth are the quantum Lyapunov exponent $\lambda_L$ defined via the exponential increase of OTOCs and the effective temperature $\beta_0$ that controls the growth rate of the Lanczos coefficients $b_n$. 

Let us define the normalised and regularised OTOC by 
\es{OTOC}{
f(t,x)={\Tr\le(y V y W(t,x) y V y W(t,x)\ri)\ov \Tr\le(y^2 V y^2 V\ri)\Tr\le(y^2 W y^2 W\ri)}\,,\qquad y=\rho_\text{th}(\beta)^{1/4}\,,
}
where $V$ and $W$ are Hermitian operators and $V=V(0,0)$ is a shorthand. In large-$N$ systems $f(t,x)$ is expected to behave as
\es{LyapGrowth}{
f(t,x)=1-{e^{\lam_L\,t}\ov N^\#}+\dots\,,\qquad \beta\ll t\ll\beta\log N\,,
}
where we assume that $x$ is held fixed.
Under some well-motivated assumptions, the exponent $\lambda_L$ was bounded in~\cite{Maldacena:2015waa} for the case of large $N$ thermal systems,
\begin{equation}
    \lambda_L \leq \frac{2\pi}{\beta}\,. 
\end{equation}

Following this, a second bound on chaos was conjectured in~\cite{Parker:2018yvk} using the asymptotic growth rate of the Lanczos coefficients. The effective temperature $1/\beta_0$ was claimed to serve as a tighter bound on chaos compared to~\cite{Maldacena:2015waa},
\es{KrylovBound}{
    \lambda_L \leq \frac{2\pi}{\beta_0} \leq \frac{2 \pi}{\beta}\,.
}
In the following we numerically verify these inequalities in finite $p$ SYK at finite temperature. In the large $p$ SYK chain we propose and analytically demonstrate a momentum-dependent generalisation of~\eqref{KrylovBound}.

The chaos bound can be refined by allowing $x$ to vary in~\eqref{LyapGrowth}~\cite{Mezei:2019dfv}. This leads to the introduction of the velocity dependent Lyapunov exponent $\lam(v)$:
\es{LyapGrowth2}{
f(t,x=vt)=1-{e^{\lam(v)\,t}\ov N^\#}+\dots\,,\qquad \beta\ll t, x\ll\beta\log N\,,
}
where 
\es{VDLEBound}{
    \lambda(v) \leq \frac{2\pi}{\beta}\le(1-{v\ov v_B}\ri)\,,
}
where $v_B$ is the butterfly velocity. In models the exponential growth usually comes from a Fourier transform
\es{LyapGrowth3}{
e^{\lam(v)\,t}=\int dp\,e^{(\lam(k)+i k v)t}\,,
}
where the Fourier transform is evaluated using saddle point with the saddle $k^*(v)$ on the imaginary $k$ axis. The bound~\eqref{VDLEBound} translates to $\lam(k^*)\leq 2\pi/\beta$.\footnote{The story is a bit more intricate: as $v$ is increased from zero, $k^*$ moves up on the imaginary axis starting from the origin. It may hit $\lam(k^*)= 2\pi/\beta$, at which point another factor in the integrand not written out in~\eqref{LyapGrowth3} has a pole. As $v$ is increased further this pole's contribution dominates over the saddle point. For the corresponding $v$'s the bound~\eqref{VDLEBound} is saturated. } Note that the more familiar Lyapunov exponent is $\lam_L=\lam(v=0)=\lam(k=0)$.

From the Fourier space two-point function, we can also define $\beta_0(k)$ as the width of the analyticity strip. We make the natural conjecture that
\es{pKrylovBound}{
    \lambda(k) \leq \frac{2\pi}{\beta_0(k)} \leq \frac{2 \pi}{\beta}\,,
}
where $k$ is taken to be real. 
Note that the conjecture does not apply to complex momenta, so in particular to the imaginary momenta that play the role of saddle point in the above discussion.
We provide a nontrivial test of this bound in the large $p$ SYK chain.

\subsection{Finite $p$ SYK}

As a demonstration of the utility of the finite temperature algorithm developed in Section~\ref{sec:scheme}, we show here that the Lyapunov exponent and effective temperature can be computed to high precision using our scheme, and as a by product, we show numerically that the bound on chaos conjectured in~\cite{Parker:2018yvk} is satisfied but not saturated in the case of finite $p$ SYK.

The bound~\eqref{KrylovBound} was proven to hold true at infinite temperature in~\cite{Parker:2018yvk} but remains a conjecture at finite $T$ for general systems. In the context of the SYK model,~\eqref{KrylovBound} was proven to be true also at finite temperature in~\cite{Gu:2021xaj}. More recently,~\cite{Chapman:2024pdw} explored this bound in the context of (deformed) SYK models. One of their main results involves using large $p$, large $\beta \mathcal{J}$ asymptotics to show that $\lambda_L < 2\pi/\beta_0$ in SYK, i.e., the inequality is strict. They further attempt to see this numerically in the case of finite $p$ but conclude that $\lambda_L \simeq 2\pi/\beta_0$ within the numerical errors of their methods (see Figure 7 in~\cite{Chapman:2024pdw}). We note that the numerical methods of~\cite{Gu:2021xaj} (see Figure 9 (b)) could show that the inequality is strict but were not precise enough to make out the functional form of the difference $\lambda_L - \frac{2\pi}{\beta_0}$.

\begin{figure}
    \centering
    \includegraphics[width=0.8\linewidth]{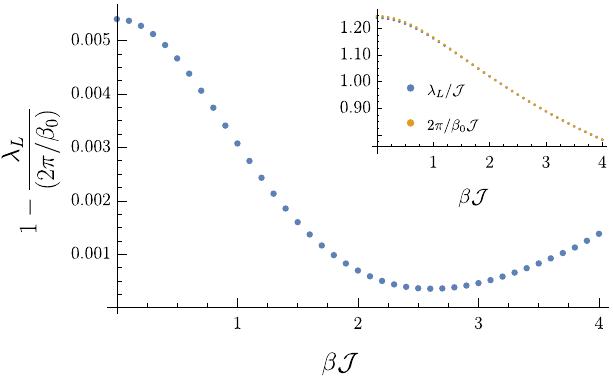}
    \caption{Comparison of the Lyapunov exponent $\lambda_L$ and effective temperature $\beta_0$ in $p= 4$ SYK in the range $0 \leq \beta \mathcal{J} \leq 4$. Our numerics allow us to show to high precision that the bound conjectured in~\cite{Parker:2018yvk}, i.e., $\lambda_L \leq 2\pi/\beta_0$ is strict in the case of finite $p$ SYK. }
    \label{fig:krylov}
\end{figure}

In our case, we are able to push further. We compute the Lyapunov exponent and effective temperature using the data contained in the moments $\mu_{2n}$, and plot the relative difference $1-\lambda_L/ \frac{2\pi}{\beta_0}$ in Figure~\ref{fig:krylov} to precision far exceeding other methods in the literature.

The effective temperature $\beta_0$ is related to the asymptotic growth rate of the Lanczos coefficients $b_n$. More precisely, the universal operator growth hypothesis asserts that the Lanczos coefficients grow asymptotically linearly 
\begin{equation*}
    b_n = \frac{\pi }{\beta_0} n + o(1) \,, \quad n \to \infty\,,
\end{equation*}
where $\beta_0$ sets the rate of growth. In fact, we already made this statement in terms of the moments $\mu_{2n}$ in~\eqref{uogh}. One can translate between $\mu_{2n}$ and $b_n$ using the recursion relation~\eqref{eq:lanczos}, and it is then straightforward to extract $\beta_0$ using a linear fit.\footnote{As noted also in~\cite{Chapman:2024pdw}, the even and odd Lanczos coefficients are staggered with respect to each other, so one must use only either the odd or even coefficients to extract $\beta_0$.}

In order to solve for the Lyapunov exponent, we follow the approach described in Section 3.6.1 of~\cite{Maldacena:2016hyu} in defining a retarded kernel for the ladder diagrams, and asserting that the OTOC is an eigenfunction of the retarded kernel with eigenvalue 1. To carry out the procedure at finite $p$,~\cite{Maldacena:2016hyu} (and~\cite{Chapman:2024pdw}) solve the real-time Schwinger-Dyson equations numerically to find $G_R(\omega)$ and $C(t)$. However, this is no longer necessary because of the finite temperature algorithm introduced in Section~\ref{sec:scheme}. Knowledge of the moments $\mu_{2n}$ gives $G_R(\omega)$ and $C(t)$ to much higher precision than numerically solving the real-time Schwinger-Dyson equations. Then we proceed as in~\cite{Maldacena:2016hyu} in solving the eigenvalue equation by representing it as a matrix equation on a discrete $\omega$ grid, and finding $\lambda_L$ by binary search. 

We plot the results of this procedure for the case of $p = 4$ SYK in Figure~\ref{fig:krylov} in the range $0 \leq \beta \mathcal{J} \leq 4$. At $\beta = 0$, the exponents already differ as noted in~\cite{Parker:2018yvk}. However, this discrepancy appears to decrease with increasing $\beta \mathcal{J}$ before reaching a minimum at $\betaJ \sim 2.5$ and increasing again. As in Section~\ref{sec:counterexamples_SYK}, we can only compute the first six moments reliably beyond $\betaJ = 4$ and we are no longer able to compute $\lambda_L$ and $\beta_0$ to sufficiently high precision.  

\subsection{Large $p$ SYK chain}

In~\cite{Choi:2020tdj} the momentum space Lyapunov exponent was computed. It is given by the simple formula
\es{ChainLam}{
\lambda(k)=\frac{2 \pi}{\beta}\, v(h(k)-1)\,,
} 
where for real $k$ we have $1\leq h(k)\leq 2$.\footnote{For inter-site coupling $\ga<1$ the minimum value of $h(k)$ is greater than $1$.} For imaginary momenta $h(k)>2$.

From the large $\om$ behaviour of the two-point function, we have found that 
\es{Chainbeta0}{
\beta_0(k)={\beta\ov v}\,,
}
independently of $k$. Thus we have found that the  momentum refined Krylov chaos bound we conjectured is obeyed in the large $p$ SYK chain
\es{pKrylovBoundSYK}{
   \frac{2 \pi}{\beta}\, v(h(k)-1)\leq \frac{2\pi}{\beta} \, v\leq \frac{2 \pi}{\beta}\,,
}
where we used that $0\leq v \leq 1$ and $1\leq h(k)\leq 2$ for physical momenta. We also see that the bound is not obeyed for imaginary momenta for which $1<(h(k)-1)$.

\section{Discussion}\label{sec:discussion}

In this paper we have extended the short time expansion method for solving the Schwinger-Dyson equations of the SYK model from~\cite{Parker:2018yvk} to finite temperature. We used the resulting Lanczos coefficients in two ways: to determine QNMs, and to study operator growth.

We found that while the QNMs form a Christmas tree shape at any fixed temperature reminiscent of AdS black holes, their temperature dependence is generically non-monotonic. Except for special examples related to symmetry, i.e., conserved densities and spontaneous symmetry breaking, this is unlike the monotonic behaviour of AdS black hole QNMs that obey the inequality~\eqref{hotfast}. 

We found that the bound on the Lyapunov exponent inspired by the universal operator growth hypothesis is satisfied as a strict inequality at finite temperature in $p=4$ SYK. We also proposed a momentum-refined bound and found that was obeyed by the large $p$ SYK chain. 

In the future, it would be interesting to extend the short time expansion to richer large $N$ systems with more complicated Schwinger-Dyson equations: to the finite $p$ SYK chain and to the melonic CFT introduced in~\cite{Murugan:2017eto}. Another direction of research could be a more systematic study of the nature of the power series in $\beta$. Our results suggest the power series is convergent with radius $\tilde{\beta}_0/2$, but it would be interesting to study more carefully the analytic structure of $C(t)$ in the complex $\beta$ plane.

In this work we have focused on the analytic structure of correlators in complex frequency. A complementary question is the analytic structure in complex time, which is a probe of the black hole singularity \cite{Fidkowski:2003nf,Ceplak:2024bja,Afkhami-Jeddi:2025wra}. At infinite temperature, the SYK model exhibits singularities in complex time which are reminiscent of the bouncing geodesics that appear in the black hole context \cite{Dodelson:2025jff}. It would be interesting to extend this analysis to finite temperatures using our method. We expect that the result is qualitatively similar to the transition from $p=4$ to $p=\infty$, where the singularities in complex time eventually disappear onto the imaginary axis, reproducing the JT gravity answer. 
\section*{Acknowledgments}
We thank Luca Delacr\'etaz, Dami\'an Galante, Vito Pellizzani, Ahmed Sheta, Julian Sonner, and Gonzalo Torroba for discussions.  We are grateful to the authors of~\cite{Chapman:2024pdw} for sharing their code.

MD’s work is supported by DOE grant DE-
SC/0007870 and the Frankel-Goldfield Research Fund. OG and
MM are supported by the ERC Consolidator Grant GeoChaos-101169611. DW acknowledges support by NSF grant PHY-2207659 and the Simons Collaboration on Celestial Holography. 

This work is funded by the European Union. Views and opinions expressed are however those of the authors only and do not necessarily reflect those of the European Union or the European Research Council Executive Agency. Neither the European Union nor the granting authority can be held responsible for them.

For the purpose of open access, the authors have applied a CC BY public copyright licence to any Author Accepted Manuscript (AAM) version arising from this submission.

\appendix

\section{Real-time Schwinger-Dyson equations in SYK at finite temperature}
\label{sec:finiteT}
Our starting point is the Schwinger-Dyson equations in imaginary time~\cite{Maldacena:2016hyu}
\begin{equation}
    \frac{1}{G(\omega_n)} = -i \omega_n - \frac{2^{p-1}\mathcal{J}^2}{p}\Sigma(\omega_n)\,,\quad \quad \Sigma(\tau) = G(\tau)^{p-1}\,.\label{eq:SDeqns}
\end{equation}
The Euclidean propagators in frequency space are defined as 
\begin{equation} \label{eq:euclidean}
    G(\omega_n) = \int_0^{\beta} d\tau\, e^{i \omega_n \tau} G(\tau)\,, \quad \quad 
    \Sigma(\omega_n) = \int_0^{\beta} d\tau\, e^{i \omega_n \tau} \Sigma(\tau)\,,
\end{equation}
where $\omega_n = 2\pi (n+1/2)/\beta$ are the Matsubara frequencies.

The first equation in~\eqref{eq:SDeqns} is trivially analytically continued as 
\begin{equation}
    \frac{1}{G(\omega_n \to -i \omega)} \equiv -\frac{1}{G_R(\omega)} = -\omega  - \frac{2^{p-1}\mathcal{J}^2}{p}\Sigma(- i \omega)\,,
\end{equation}
where we have made use of the spectral representation (see~\cite{Meyer_2011} for example) in analytically continuing the Matsubara correlator to the retarded Green's function in frequency space. Then the non-trivial part is analytically continuing $\Sigma(\omega_n)$.

To do this, note that it is possible to write the Euclidean correlator for $\tau > 0$ as\footnote{The factor of 2 arises because we choose the normalisation where $\mu_{0,0}=1$ for $C(t)$.}
\begin{equation}
    G(\tau) = \int_{-\infty}^{\infty} \frac{d \omega}{2 \pi}\, e^{- \omega \left(\tau-\frac{\beta}{2}\right)} \frac{C(\omega)}{2}\,. \label{eq:2.4}
\end{equation}
Then we have
\begin{equation}
    \Sigma (\omega_n) = \int_0^{\beta} d\tau\, e^{i \omega_n \tau - \sum_j \omega_j \left(\tau-\frac{\beta}{2}\right)} \int \prod_{j = 1}^{p-1}\left(\frac{d\omega_j}{2 \pi}\,\frac{C(\omega_j)}{2} \right)\,.
\end{equation}
Doing the $\tau$ integral and Fourier transforming $C(\omega)$, we find
\begin{equation} \label{eq:interstep}
    \Sigma (\omega_n) = 2  \int \left(\prod_{i = 1}^{p-1} \frac{d \omega_i}{2 \pi} \right)\int \left( \prod_{j = 1}^{p-1} d t_j  \, \frac{C(t_j)}{2} \right) e^{i \sum_i \omega_i t_i} \frac{\cosh \left(\frac{\beta \sum_i \omega_i}{2}\right)}{\sum_i \omega_i-i \omega_n }\,.
\end{equation}

Let us now specialise to $p =4$ and make the following change of variables,
\es{newoms}{
    \omega_+ &=  \omega_1 + \omega_2 + \omega_3\,,\\
    \omega_{12} &= \omega_1 - \omega_2\, ,\\
    \omega_{23} &= \omega_2 - \omega_3\,.
}
Then we find
\es{tripleint}{
    \Sigma (\omega_n) & = \frac{1}{4}\int   \frac{d \omega_+\, d\omega_{12}\, d \omega_{23}}{3 (2 \pi)^3}  \int \left( \prod_{j = 1}^{3} dt_j\,  C(t_j) \right)  \, \frac{\cosh \left(\frac{\beta \omega_+}{2}\right)}{ \omega_+-i \omega_n } \\
    & \times\exp\left( \frac{i}{3} (\omega_{23} (t_1 + t_2 - 2t_3) + \omega_{12} (2t_1 - t_2 -t_3) + \omega_+(t_1 + t_2 + t_3)) \right)\,.
}
The $\omega_{12}$ and $\omega_{23}$ integrals can be done trivially, and the delta functions enforce $t_1=t_2=t_3$. The integral then reduces to
\begin{equation}
    \Sigma(\omega_n) = \frac{1}{4} \int \frac{d \omega}{2 \pi}  \,\frac{\cosh \left(\frac{\beta \omega}{2}\right) }{\omega-i \omega_n} \int d t\, e^{i \omega t} C(t)^3\,.
\end{equation}
\indent For general $p$, one can change variables to the pairwise differences $\omega_{i,i+1}$ along with the total energy $\omega_+=\sum \omega_i $. The general result is
\begin{equation}
    \Sigma(\omega_n) = 2^{2-p} \int \frac{d \omega}{2 \pi}  \frac{\cosh \left(\frac{\beta \omega}{2}\right)}{\omega-i \omega_n } \int d t\, e^{i \omega t} C(t)^{p-1}\,.\label{eq:2.16}
\end{equation}

\section{Monotonicity from Bohr-Sommerfeld quantization} \label{app:BSproof}

The integral in~\eqref{BSquant} can be performed explicitly, and gives a hypergeometric function
\es{BSquant2}{
\frac{\pi\left(\Delta-\frac{d}{2}+1+2n\right)}{2}&=\sqrt{k^2-\omega_n^2}\int_{r_T}^{\infty}dr\,\frac{\sqrt{1-\left(\frac{r_T}{r}\right)^d}}{r^2-\mu r^{2-d}}\\
&=-{ik\ov \mu^{{1\ov d}}}\,{\sqrt{\pi}\,\Ga\le(1+{1\ov d}\ri) \le(1-u^2\ri)^{\frac12+{1\ov d}} {}_2F_1\le(1,{1\ov d},\frac32+{1\ov d},1-u^2\ri)\ov  2 \Ga\le(\frac32+{1\ov d}\ri)}\,,
}
where we assumed $k>0$ and where $u=\omega_n(k)/k$. We take the $\mu$ derivative of this equation and use hypergeometric identities to find that the result can be expressed solely in terms of the hypergeometric function appearing in~\eqref{BSquant2}. We then express this hypergeometric function using the left hand side of~\eqref{BSquant2} to obtain the following simple formula:
\es{omder}{
&\le[\frac{\pi\left(\Delta-\frac{d}{2}+1+2n\right)}{2}+{ik\ov \mu^{{1\ov d}}}\,{\sqrt{\pi}\,\Ga\le(1+{1\ov d}\ri) \le(1-u^2\ri)^{-\frac12+{1\ov d}} \ov   \Ga\le(\frac12+{1\ov d}\ri)}\ri]{d\om_n(k)\ov d\mu}\\
&=\frac{\pi\left(\Delta-\frac{d}{2}+1+2n\right)}{2 d\mu}\om_n(k)\,.
}
From this equation, using that $\om_n(k)$ is in the fourth quadrant and that $k>0$, we will now show that ${d\om_n(k)\ov d\mu}$ is in the lower half plane, which implies monotonicity. Since $\left(\Delta-\frac{d}{2}+1+2n\right)>0$ and $\om_n(k)$ is in the fourth quadrant, the right hand side is in the fourth quadrant. Since $u$ is in the fourth quadrant, $\le(1-u^2\ri)^{-\frac12+{1\ov d}}$ is in the fourth quadrant and multiplying it with $i$ and the other positive factors brings this term into the first quadrant. Hence the square bracket on the left hand side is in the first quadrant. Dividing by it to express ${d\om_n(k)\ov d\mu}$ we get a fraction of the form (fourth quadrant)$/$(first quadrant) $\in$ LHP, which completes the proof.

\bibliographystyle{JHEP}
\bibliography{mybib}
  
\end{document}